 \documentclass[final,5p,times,twocolumn]{elsarticle}

\usepackage{amssymb}
\usepackage{amsmath}
\usepackage{lineno}

\usepackage{url}
\usepackage{comment}

\usepackage{caption}
\usepackage{subcaption}
\usepackage{widetext}
\usepackage{float}
\usepackage{placeins}
\usepackage{adjustbox}

\journal{Acta Astronautica}

\begin{document}

\date{}

\begin{frontmatter}



\title{Deflective Sunshades: conceptual design and origami-inspired folding strategy}


\author[actaffiliation]{Benedetta Marazzato \fnref{equal}} %% Author name
\author[actaffiliation]{Jonas Seiler \fnref{equal}}
\author[actaffiliation]{Harry Holt \fnref{equal}}
\author[actaffiliation]{Dolf Huybrechts \fnref{equal}}
\author[actaffiliation]{Richard Murchie \fnref{equal}}
\author[actaffiliation]{Leone Costi \fnref{equal}}
\author[actaffiliation]{Jai Grover \fnref{equal}}
\author[actaffiliation]{Dario Izzo \fnref{equal}\corref{cor1}}

\fntext[equal]{All authors contributed equally to this work.}

\cortext[cor1]{Corresponding author at: European Space Agency, ESTEC, 2201AZ, Noordwijk, Netherlands}
\ead{dario.izzo@esa.int}

\affiliation[actaffiliation]{organization={Advanced Concepts Team, European Space Agency},%Department and Organization
            addressline={Keplerlaan 1}, 
            city={Noordwijk},
            %%postcode={2201 AZ}, 
            %%state={},
            country={The Netherlands}}

%% Abstract
\begin{abstract}
Space-based solar radiation management has been proposed as a direct means of modulating Earth’s radiative balance by reducing incident solar flux. A prominent concept consists of large sunshades operating near the Sun--Earth $L_1$ region. However, feasibility is constrained by a fundamental lower bound on the required mass, arising from the coupled requirements of achieving sufficient insolation reduction while maintaining dynamical equilibrium under solar radiation pressure (SRP). We introduce a class of sunshades, termed deflective, designed to shape the effective SRP through macroscopic surface geometry. By employing inclined reflective elements, these systems redirect incident radiation to simultaneously control flux attenuation and the resulting momentum exchange, enabling the use of conventional high-reflectivity materials (e.g., aluminium films). The concept admits multiple geometrical realizations, including conical, pyramidal, and louvered (venetian-blind) configurations. A conical configuration is then studied in more detail as a representative implementation of this broader design principle. To improve its packaging efficiency and scalability, we consider a distributed constellation of units and propose a flat-folding strategy based on a Miura--Ori pattern adapted to the conical geometry finding its optimal configuration parameters resulting in a favourable geometry.
\end{abstract}


%%\item Research highlight 1
%%\item Research highlight 2

%% Keywords
\begin{keyword}
Sunshade \sep Geoengineering \sep Origami


\end{keyword}

\end{frontmatter}

%\linenumbers


\section{\label{sec:intro}Introduction}

%Atmospheric CO2 concentrations and global temperatures are rising rapidly.
\noindent According to NASA’s Goddard Institute for Space Studies, atmospheric $\mathrm{CO_2}$ concentrations increased from approximately 365 ppm in 2000 to more than 425 ppm in 2025 \cite{NASA_carbon_dioxide}. This sustained rise has contributed to an increase in global mean temperature, with 2025 reaching approximately 1.2$^\circ$C above the pre-industrial level \cite{NASA_global_temperature}. This value indicates that the climate system is already approaching the temperature thresholds established by the Paris Agreement, which aims to limit global warming to well below 2$^\circ$C above pre-industrial levels, while pursuing efforts to restrict the increase to 1.5$^\circ$C \cite{ParisAgreement}. This context has motivated an increasing interest in large-scale climate intervention strategies, commonly referred to as geoengineering \cite{geo}. Among these, solar radiation management aims to modify Earth’s radiative balance by reducing incoming solar radiation or altering outgoing terrestrial infrared radiation.

Within this framework, space-based sunshades have been investigated as a possible solar radiation management strategy \cite{firstsunshade,mcinnes}. These concepts generally involve the deployment of high area-to-mass ratio structures near the Sun-Earth $L_1$ region to partially attenuate the incoming solar flux. Among the architectures proposed in the literature, McInnes' model \cite{mcinnes} represents one of the main reference cases for space-based sunshades. In their work, the sunshade is modelled either as a single large occulting disc or as a constellation of smaller disc elements. Their analysis shows that the optimum location of such a system lies sunward of the classical $L_1$ Lagrange point and highlighted the primary constraint of the concept: the substantial system mass required to achieve the necessary reduction in solar irradiance. 
To mitigate this constraint, McInnes suggested the use of lunar or in-situ resources, including near-Earth asteroid material, with small M-type asteroids representing a potential source due to their abundance in nickel-iron materials and carbon. 

Several alternative sunshade concepts have been proposed with the aim of reducing the overall system mass. These approaches act on different design variables, including material properties, geometric configuration, and system architecture.

One highly challenging concept based on extraterrestrial resources consists of placing clouds of dust grains at the stable Earth-Moon triangular Lagrange points \cite{struck}, $L_4$ and $L_5$. In this scheme, approximately $2 \cdot 10^{11}$ tonnes of lunar or cometary dust would be introduced into the Earth-Moon system to form occulting clouds capable of reducing solar insolation. A more recent variant of the dust-cloud approach is the JPL DimSun mission concept \cite{dimsun}, which proposes transporting small bodies to the Sun-Earth $L_1$ region and pulverizing them into micron-scale regolith particles to form an occulting cloud. The cloud would be confined using multiple spacecraft that apply repulsive forces through ion-thruster-based interactions, but its feasibility is limited by the need to transport, pulverize, and actively confine  dust.

However the need for extensive use of extraterrestrial resources and the associated manufacturing challenges make these architectures infeasible with current technology. 

A different strategy was proposed by Angel \cite{angel}, based on transparent, refractive thin screens, deployed as a large cloud of meter-scale autonomous spacecraft. The proposed refractive screen consisted of a silicon nitride core with silica-based coatings, incorporating holes to reduce mass while maintaining optical performance. Building on the same underlying principle of refractive solar flux deflection, Borgue et al. \cite{transparent_occulters} introduce a nearly zero-radiation pressure sunshade based on transparent refractive membranes fabricated from ultra-thin, low refractive index $\mathrm{SiO_2}$ nanotube films with a sawtooth profile. This configuration is reported to be $10^2$--$10^3$ times lighter than the lightest existing sunshade concepts; however, its practical implementation remains constrained by unresolved challenges in the manufacturing, integration, and validation of ultra-thin structured membranes at sunshade scale.

Finally, Szapudi proposes a tethered sun-shield concept \cite{tether}, in which solar radiation pressure is balanced at a modified Lagrange point by connecting the shield to a counterweight positioned sunward. Although this configuration reduces the minimum mass compared to the McInnes shield, the requirement for extremely long tethers, of the order of $1$--$3\,\mathrm{Mkm}$, represents a major viability constraint. 

Given the limitations and feasibility constraints of the architectures discussed above, we propose a deflective sunshade made of aluminium which reduces the mass threshold by deflecting incident sunlight, without relying on advanced materials. The proposed approach therefore introduces a trade-off between the mass reduction enabled by geometric modification and the engineering feasibility associated with the use of aluminium, a material widely used in space applications and abundant on Earth. This architecture may be implemented through different geometrical configurations, including conical, pyramidal, or venetian-blind-like arrangements. The underlying principle is analogous to the light-deflection mechanism proposed by Borgue et al. \cite{transparent_occulters} for sawtooth transparent membranes; however, in the present case, deflection is achieved through the external geometry of the sunshade rather than through the internal structuring of the membrane material.

A second novelty of the proposed concept is the introduction of a flat-folding strategy to improve packaging efficiency and enable scalable deployment. Origami-inspired designs have become a promising solution for space applications, offering high deployment ratios and adaptable geometries \cite{ravichand2024origami}. We propose a strategy based on a generalised Miura-Ori pattern \cite{sharma2021folding} tailored to the conical geometry. Miura-Ori patterns have been widely investigated for deployable space structures and offer several favourable properties, including a single degree of freedom, rigidity, and flat foldability \cite{misseroni2024origami} \cite{fang2017dynamics}. Deployment and stabilization by spinning have also been studied as a viable alternative to large control systems \cite{yuan2022comparison} \cite{gardsback2007design}. The proposed design should be interpreted not as a definitive or uniquely optimal solution, nor as evidence that such a sunshade could be realised in the near term, but rather as a step towards improving the feasibility of space-based sunshade concepts previously proposed in the literature.

This paper is organised as follows. In Sec.(\ref{sec:theory}) we discuss the necessary background theory for the preliminary design of a sunshade. Sec.(\ref{sec:architecture}) introduces the deflective architecture adopted in this paper, comparing it with previous sunshade design. A conical configuration is also introduced as a representative deflective sunshade. Sec.(\ref{sec:coneorigami}) presents a generalised Miura-ori pattern applied to conical shapes, with an unfolding mechanism based on centrifugal force. In Sec.(\ref{sec:system}) we discuss some preliminary system-level considerations for a constellation of cones. Finally, the conclusions are drawn in Sec.(\ref{sec:conclusions}).

%%%%%%%%%%%%%%%%%%%%%%%%%%%%%%%%%%%%%%%%%%%%%%%%%%%%%%%%%%%%%%%%%%%%%%%%%%%%%
% \clearpage

\section{\label{sec:theory}Background theory}

\noindent  According to the energy-balance model adopted by McInnes \cite{mcinnes}, compensating for a projected 2 K increase in Earth’s temperature would require an approximately 1.7\% reduction in incoming solar insolation, $\delta Q/Q$. This estimate is based on a simplified climate model and therefore should not be interpreted as an exact requirement. More detailed analyses of the climatic response were carried out by Sanchez and McInnes \cite{sanchez}, using a globally resolved energy-balance model to investigate both global and regional temperature effects. However, the 1.7\% attenuation provides a useful baseline value for preliminary analysis. For this reason, it is adopted here as the reference target for the proposed sunshade system. Since the present model is parametric, different attenuation targets can also be evaluated without changing the overall methodology.

\begin{figure}[h!]
    \centering
    \includegraphics[width=0.9\columnwidth]{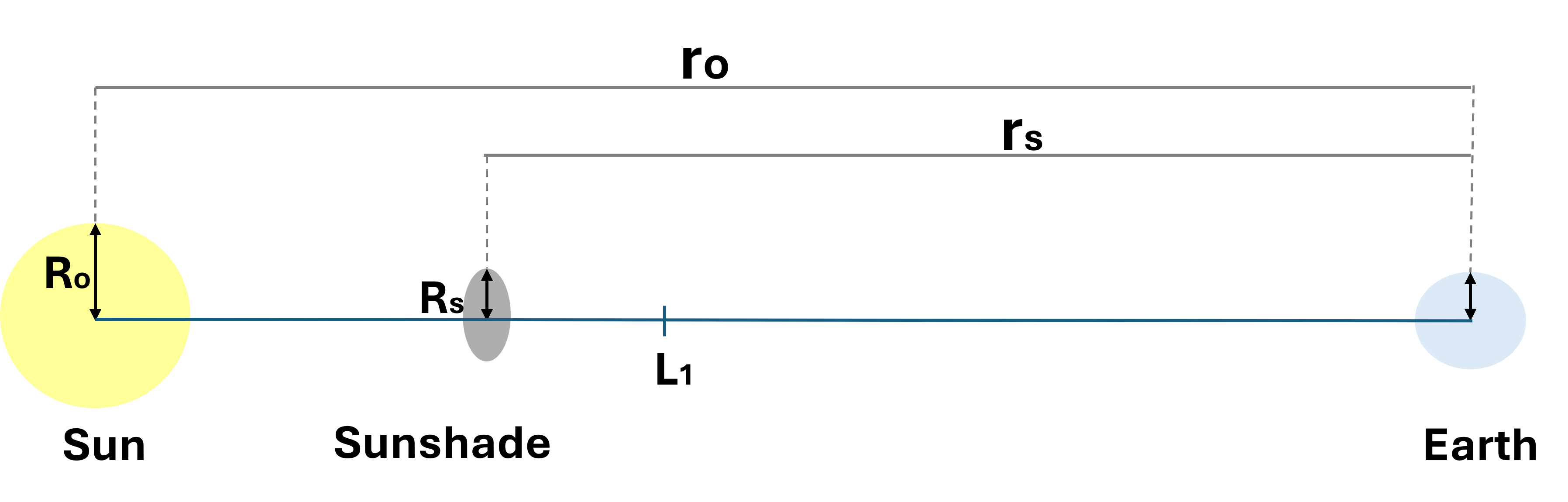}
    \caption{One-dimensional model of the system, showing the sunshade positioned along the Sun–Earth line near the artificial L1 equilibrium point.}
    \label{fig:def1}
\end{figure}

As first shown by McInnes \cite{mcinnes}, the design of a sunshade, namely its mass, size, and optimal orbital configuration, can be determined by enforcing two coupled conditions.
The first is a geometric requirement: the shadow cast by the sunshade must yield the desired fractional reduction in solar irradiance on Earth, $\delta Q/Q$. This condition defines the effective radius of the structure as
\begin{equation}
\label{eq:radius}
R_s= R_\circ \left(\frac{r_s}{r_\circ}\right)\left(\frac{\delta Q}{Q}\right)^{1/2} ,
\end{equation}
where $R_\circ=6.963 \cdot 10^8$ m is the radius of the Sun and $r_\circ=1.51 \cdot 10^{11}$ m is the Sun-Earth distance (Fig.\eqref{fig:def1}), $R_s$ is the radius of the occulting disc and $r_s$ is its distance from Earth.

The second is a dynamical requirement prescribing the additional non-Keplerian acceleration necessary for a sunshade located at distance $r_s$ to remain on a stable circular orbit under the combined action of solar and terrestrial gravity and the centrifugal term associated with the rotating frame. This acceleration is given by
\begin{equation}
\label{eq:1Ddynamics}
a_g(r_s) = -\left(\frac{G M_E}{r_s^2} - \frac{G M_\circ}{(r_\circ - r_s)^2} + \omega^2(r_\circ - r_s) \right),
\end{equation}
where $\omega=\sqrt{G M_\circ/r_\circ^3}$ is the angular velocity of the Earth's circular orbit around the Sun. For $a_g=0$, the classical Lagrange equilibrium point is recovered.

Assuming solar radiation pressure is the only source of non-Keplerian acceleration, the required acceleration can also be written as
\begin{equation}
\label{eq:SRP}
a_s(r_s) = \frac{2\kappa P_E A_s}{M_s}\left(\frac{r_\circ}{r_\circ - r_s}\right)^2,
\end{equation}
where $A_s=\pi R_s^2$ represents the sunshade area projected into the Sun-Earth direction, $M_s$ is its total mass and $P_E=4.56 \cdot 10^{-6}$ N/m$^2$ is the solar radiation pressure at 1 AU and $\kappa$ defines an effective momentum-transfer coefficient that accounts for the optical properties of the sunshade.
\\
Imposing the condition $a_g = a_s$ and combining Eqs.~\eqref{eq:1Ddynamics}, \eqref{eq:SRP}, and \eqref{eq:radius}, the required sunshade mass can be related to the prescribed change in insolation $\delta Q/Q$:
\begin{equation}
\label{eq:Mass}
\begin{aligned}
M_s(r_s) &= \frac{2\kappa P_E A_s}{a_g(r_s)}\left(\frac{r_\circ}{r_\circ - r_s}\right)^2= \\ &
= 2 \pi \kappa P_{E} R_\circ^2
\left( \frac{\delta Q}{Q} \right)
\left( \frac{r_s}{r_\circ - r_s} \right)^2
\frac{1}{a_g(r_s)}.
\end{aligned}
\end{equation}
which we also write as $M_s(r_s) = \kappa \overline M_s(r_s)$ to highlight the simple dependence on $\kappa$.
\subsection{\label{preliminary_design}Preliminary design of sunshades}

\noindent Eq.~\eqref{eq:Mass} (as derived by McInnes \cite{mcinnes}) shows that, for a prescribed fractional reduction in insolation, the sunshade mass is determined by the balance between a geometric constraint and a dynamical equilibrium condition. As the sunshade moves closer to the Sun, the incident solar radiation pressure increases without bound, while the characteristic size of the sunshade remains finite because it is fixed by the imposed occultation requirement. Therefore, the mass required to produce the prescribed finite equilibrium acceleration must also increase without bound. A similar divergence occurs as the equilibrium approaches the classical Lagrange point, but for a different reason: the required non-Keplerian acceleration tends to zero. Consequently, $M_s(r_s)$ is unbounded at the limiting configurations and will therefore attain a finite relative minimum $M_s^*$ at some intermediate value $r^*_s$.

Hence, Eq.~\eqref{eq:Mass} provides the mandatory starting point for any preliminary sunshade design: the orbital location $r_s^*$ where $M_s(r_s)$ has its relative minima is fixed, while the effective radius $R_s$ is prescribed by $\delta Q/Q$. The associated mass $M_s(r_s) = \kappa \overline M_s(r_s)$ depends linearly on the only design parameter, $\kappa$, that accounts for the optical properties of the shade. 
The reference mass $\overline M_s(r_s^*)$, evaluated at $\kappa=1$, is a thus an important quantity for quickly assessing the engineering feasibility of any sunshade concept realising some $\kappa$.

As a numerical example, consider a target insolation reduction of $\delta Q/Q = 1.7\%$. In this case, the effective radius is approximately $1430~\mathrm{km}$, the optimal location is $r_s^* = 2.36\cdot 10^{6}~\mathrm{km}$. 
The corresponding mass $\overline M_s(r_s^*)$ is $2.8\cdot 10^{11}~\mathrm{kg}$. 
By contrast, for $\kappa\approx 0.17$, which corresponds to a non‑reflective black disc, the total mass scales down linearly to $4.8\cdot 10^{10}~\mathrm{kg}$, while the effective radius and the optimal location $r_s^*$ remain unchanged. 
%\ref{appendix:analytic_rs} derives an analytic approximation of the optimal location and demonstrates how $\kappa$ does not affect the optimal location in this model. 

This illustrates how, once $\overline M_s(r_s^*)$ is computed, the design can be easily adjusted for any other value of $\kappa$ by simple linear rescalings.

This simple and elegant description suggests the design of sunshade concepts to be focused primarily on the optical property $\kappa$ appearing in Eq.~(\ref{eq:SRP}). Independently of the underlying mechanism, whether achieved through nanostructured materials or macroscopic geometry, $\kappa$ captures the net momentum-transfer efficiency of any sunshade (with the exception of the tethered-sunshade concept \cite{tether}): it encodes how incident photons are reflected, absorbed, or redirected, and therefore how much radiation pressure is generated. Designing an effective preliminary sunshade concept therefore reduces to engineering a favourable value of $\kappa$, while $r_s^*$ the value $\overline M_s(r_s^*)$ and the effective radius $R_s$ are determined independently by the attenuation $\left(\frac{\delta Q}{Q}\right)$ requirement.

\section{\label{sec:architecture}Deflective sunshades}

\noindent
We now study a class of sunshades, here termed \emph{deflective}, whose purpose is to engineer a small value for $\kappa$ through the macroscopic design of the surface geometry. 
In contrast to conventional occulting concepts, which rely primarily on blocking the incoming solar flux, a deflective sunshade is designed so that the incident photons are only deflected the necessary amount to miss the Earth via some well thought macroscopic arrangement of the surface geometry. 
The resulting radiation-pressure force can still be described within the framework introduced in Sec.\eqref{sec:theory}, provided that the corresponding value of $\kappa$ is introduced. 
For a surface of physical illuminated area $A$ tilted by an angle $\alpha$ with respect to the plane orthogonal to the incident solar rays, we define an effective momentum-transfer coefficient $\kappa$ starting from the radial solar-radiation-pressure force model used also in Eq (\ref{eq:SRP}):
\begin{equation}
\label{eq:kappa_def}
F_{srp} = 2 \kappa P_E\left(\frac{r_o}{r_o-r_s}\right)^2 A_s,
\end{equation}
where $F_{srp}$ is the magnitude of the force due to the solar radiation pressure in the direction from the Sun to Earth, and $A_s = A \sin\alpha$. Following the approach of Farres \cite{propellantless}, the solar-radiation-pressure force acting on a surface element may be decomposed into three contributions associated with absorption, specular reflection, and diffuse reflection. Let $f_a$, $f_s$, and $f_d$ denote the corresponding fractions of the incident radiation, with
\begin{equation*}
f_a + f_s + f_d = 1, \quad f_a\ge0, f_s\ge 0, f_d\ge 0
\end{equation*}
Furthermore, let us allow our flat element to have different coatings on the two sides, so that $\varepsilon_F$ and $\varepsilon_B$ are the emissivities of the front and rear sides associated with Lambertian thermal re-emission. It is then possible (see \ref{appendix:force_derivation}), to find the following expression for the momentum transfer coefficient:
\begin{equation}
\kappa(\alpha)
= \frac{1}{2}\left[
f_a+2 f_s \sin^2\alpha+f_d\left(1+\frac{2}{3}\sin\alpha\right)+\frac{2}{3}f_a\sin\alpha\frac{\varepsilon_F-\varepsilon_B}{\varepsilon_F+\varepsilon_B} \right]
\label{eq:kappa_eff}
\end{equation}

%\begin{figure}[tb]
    %\centering
   % \includegraphics[width=1\linewidth]{Figures/conediagram.png}
   % \caption{Forces acting on a deflective sunshade: ($F_{\mathrm{Sun}}$) and ($F_{\mathrm{Earth}}$) are the gravitational attractions of the Sun and Earth, respectively, while ($F_{\mathrm{centrifugal}}$) represents the centrifugal term in the rotating frame. ($F_{\mathrm{incident}}$) is the force associated with the incoming solar radiation. The reflected radiation pressure contribution is decomposed into radial and tangential components, ($F_r$) and ($F_t$), respectively.}
   % \label{fig:diagcone}
%\end{figure}

\noindent
Eq.~\eqref{eq:kappa_eff} provides the desired link between the macroscopic architecture of a deflective sunshade and the preliminary design framework introduced in Sec.\eqref{sec:theory}. In particular, many sunshade concepts may be interpreted as a physical realization of this equivalent effective momentum-transfer coefficient $\kappa$, regardless of whether this is achieved through a conical surface, a pyramidal arrangement, a Venetian-blind configuration, or any other system of inclined reflecting elements assuming no secondary reflections are significant.

\subsection{Making $\kappa$ small}
\label{sec:smallkappa}
Lowering the value of $\kappa$ is the primary aim of most sunshade and geoengineering designs. Several distinct strategies can be used to reduce $\kappa$. A number of past works \cite{mcinnes, Fuglesang, sanchez}, for example, studied surfaces perpendicular to the incoming radiation $\left(\alpha=90^\circ\right)$ and assumed the absence of diffuse reflection (i.e. $f_d=0$, $f_a = 1 - f_s$); in these cases Eq.~\eqref{eq:kappa_eff} reduces to
\begin{equation}
\kappa
=\frac 12 \left[
1+f_s + \frac{2}{3}(1-f_s)\frac{\varepsilon_F-\varepsilon_B}{\varepsilon_F+\varepsilon_B}\right].
\label{eq:kappa_perp}
\end{equation}
Whenever  $f_s=0$, $\varepsilon_B=1$ and $\varepsilon_F=0$, corresponding to a fully absorbing disc able to emit most of the radiation from its back side, one obtains $\kappa=\frac 16\approx0.17$. This number, first derived by McInnes \cite{mcinnes} and often used in the past as a benchmark for concept proposals and architectural studies on sunshades, does not correspond to a specific material and should rather be interpreted as a lower bound on what may be achievable by exploiting differential thermal emission from the two plate sides through a suitable coating design.

A different strategy pursued here is instead to assume a perfectly reflective surface, $f_a=f_d=0$, $f_s=1$, in which case Eq.~\eqref{eq:kappa_eff} gives $\kappa = \sin^2\alpha$; hence, the aim is to choose the angle $\alpha$ to be as small as possible. The minimum angle required to deflect incident radiation beyond the Earth for an $\alpha$-inclined surface placed at an offset $r$ from the Earth--Sun line can be evaluated as
\begin{equation*}
    \phi(r) = \frac{1}{2}\tan^{-1}\left(\frac{R_e-r}{r_s}\right),
\end{equation*}
where $R_e$ is the radius of the Earth and $r_s$ is the distance to the center of the Earth. Since $\phi(0) \geq \phi(r)$ for all $r$, the angle evaluated at $r=0$ is sufficient to ensure that the reflected radiation is redirected away from the Earth regardless of the distance of the surface element from the Earth--Sun line. For $r_s \approx 2 \cdot 10^{6}~\mathrm{km}$, $R_e = 6378~\mathrm{km}$, and $r=0$, the required angle is $\phi(0) \simeq 0.091^\circ$, corresponding to a value of $\kappa = 2.5\cdot10^{-6}$, which suggests a potentially dramatic reduction in launch mass. Physically, this configuration means that photons hitting the sunshade are deflected only by the amount needed to miss the Earth, while generating very little force on the structure; as a consequence, the structure may be much lighter while still producing the small non-Keplerian acceleration required at equilibrium $(M_s^* = \kappa \overline M_s^*)$.

Per se this result is not surprising as the overall path taken by the photons is similar to that promised by the use of transparent refractive materials \cite{transparent_occulters} which also, if left free of engineering constraints, promise a similar reduction in $\kappa$. However, the practical feasibility of Borgue and Hein’s approach remains limited by significant technological uncertainties. In particular, at the required scale, the sawtooth structure proposed for ultra-thin free-standing $\mathrm{SiO_2}$ nanotube films would operate in the sub-wavelength regime for visible light, where ray-optics based microprism modelling is no longer applicable. In addition, although nanoimprint lithography represents a promising fabrication path, its applicability to the large scale manufacturing of such ultra-thin, structured membranes has not yet been demonstrated. Their second design is a  polymeric thin-film membrane with a sawtooth profile and $\mathrm{SiO_2}$  anti-reflection coating, which is  more technologically mature, as its key components have been demonstrated separately. However, its feasibility remains limited by the integration of these components. Even high-throughput processes such as roll-to-roll embossing would still imply very long production times, while the optical and space environment compatibility of the required coatings and photopolymers remains to be fully assessed. Similarly to these cases, where the feasibility of engineering the computed $\kappa$ lies on the details of material fabrication and availability, the \lq\lq macroscopic\rq\rq\ alternative studied here hits a rather high bound due to engineering limits.

\subsection{Material feasibility constraints}
Two main issues drastically limit the possibility of achieving the theoretical lower bound $\kappa = 2.5\cdot10^{-6}$. First, as the angle decreases, the illuminated structural area $A = \frac{A_s}{\sin\alpha}$ grows rapidly and soon becomes impractically large. Consequently, practical engineering limitations substantially constrain the achievable configuration. Additionally, as the structure length increases, the assumption of a constant solar radiation pressure, $P_{\mathrm{srp}}$, may no longer hold, and the simplified model may become invalid.

Second, perfectly specularly reflective materials do not exist in practice: any material has some diffusive or absorptive component, depending on surface roughness, coating quality, and wavelength-dependent optical properties. This means that the idealised performance predicted for a purely specular surface can only be approached asymptotically, not achieved in reality.

As for the second limitation, it is possible to use very reflective materials that can achieve optical properties close to those of an ideal specular surface. Aluminium, silver, Kapton and Mylar are among the most common reflective materials in space applications due to their high optical reflectivity, availability, and well established fabrication processes. Aluminium, in particular, is one of the most advantageous material for its abundance and low production cost, making it one of the most versatile space materials used for both structural and optical purposes \cite{ESA2023aluminium}.  Other noble metals such as gold, platinum, and iridium are also utilised in specific optical or thermal control coatings for their stability and resistance to atomic oxygen erosion~\cite{NASA_Materials2018}. However, their use is limited mainly by their high cost, limited terrestrial abundance, and the complexity of deposition processes required to obtain optically uniform coatings. Spectralon \cite{spectralon} is also a highly reflective material, commonly used for the calibration of optical instruments. However, its reflection is predominantly diffuse rather than purely specular, making it unsuitable for the present application.

For the reasons discussed above, aluminium emerges as an excellent candidate for sunshade. The optical properties of a material are generally wavelength dependent. In the visible range, smooth aluminium surfaces typically exhibit a high total reflectance, often close to or above $90\%$ for clean polished surfaces, although the split between specular and diffuse reflection strongly depends on surface roughness, oxidation, and manufacturing process. As a representative aluminised reflective surface, we therefore assume: $f_s \approx  0.87$, $f_d \approx 0.03$, $f_a \approx 0.10$ \cite{90aluminium,AL_values, designsolarsail}. Since the two sides are assumed to have the same material, coating, and surface finish, the front- and back-side emissivities are taken to be equal, i.e. ($\varepsilon_F$ = $\varepsilon_B$).

\subsection{\label{sec:Sizingofthecone}Sizing of a deflective sunshade}

As shown in Sec.\eqref{sec:smallkappa}, reducing $\alpha$ decreases the minimum required mass. The structure is therefore sized by considering the smallest angle that can be practically achieved. This lower bound cannot be chosen independently of the material properties, since smaller angles correspond to lower required areal densities. The areal density can be expressed as the ratio between the membrane mass and its surface area, $\sigma=M_s/A_s$. For a membrane of density $\rho$ and thickness $t$, it can also be written as $\sigma=\rho t$. Therefore, the minimum attainable thickness imposes a lower bound on the areal density, $\sigma_{\min}=\rho t_{\min}$, and consequently constrains the smallest feasible angle. In physical terms, reducing the surface inclination decreases the deflective effectiveness of each unit area. As a result, a larger total surface area is required to provide the same effective reduction of the incoming solar flux. This geometric penalty can only be compensated by reducing the membrane areal density, i.e. by adopting thinner membranes. Therefore, for a given optical performance, lower inclination angles become advantageous only if sufficiently low membrane thicknesses can be achieved.

The technological feasibility of such low areal densities is closely related to the development of extremely thin, large-area reflective membranes. This topic has been extensively investigated within the solar sail community, where similar requirements arise in terms of membrane thickness, packaging volume, and deployment. For instance, JAXA’s \textit{IKAROS} \cite{ikaros} mission successfully demonstrated the in-space deployment of a 
7 $\mu \mathrm{m}$ polyimide sail with an area of 200 $\mathrm{m^2}$. 
Subsequent CubeSat solar sail demonstrations, including NanoSail-D2 and LightSail-A, employed aluminium coated polymer membranes, such as CP1 and Mylar \cite{nanosail, Lightsail2}. More recently, NASA’s Advanced Composite Solar Sail System (ACS3) demonstrated the deployment of a $2~\mu \mathrm{m}$ aluminised PEN sail in low Earth orbit, supported by deployable composite booms \cite{acs3}. According to Gong and McDonald \cite{revsolarsail}, one of the lightest proposed solar sail designs used a $0.9\,\mu \mathrm{m} $ Mylar film with aluminium and chromium coatings, giving an areal density of $2\,\mathrm{g/m^2}$. 
These examples show that extremely thin reflective membranes have already been considered and, in some cases, demonstrated in deployable space systems. 

Since aluminium is selected here as the reference reflective material, the relevant technological limit is associated with the minimum attainable thickness of aluminium-based films. Aluminium foils with thicknesses in the micrometre range are already produced industrially, with commercial alufoil generally ranging from about 6 to 200~$\mu$m \cite{alufoil}. Ultra-thin aluminium foils are also reported in the range of approximately 5--10~$\mu$m for packaging and laminate applications \cite{5_6aluminium}. Thinner aluminium foils are also commercially available as ultra-thin or nanoscale products, reaching the sub-micrometre to micrometre range \cite{1micronAL,nanoAL}. In addition, recent experimental work has demonstrated the fabrication of free-standing ultrathin aluminium foils with thicknesses of approximately $2~\mu$m using accumulative pack rolling, showing that few-micrometre aluminium foils can be obtained through mechanical rolling routes, although surface roughening, ridging, and plastic instability become relevant limitations at these scales \cite{2althickness}. 
Therefore, in this work a thickness of $t_{\min} = 1 ~\mu\mathrm{m}$ is adopted as a reference value for the following analyses. Considering this value and assuming an aluminium density of $\rho \simeq 2700~\mathrm{kg/ m^{3}}$, the minimum areal density  can be estimated, together with the associated  $\alpha$-angle, $k$ and sunshade mass $M_S$: the results, also reported in Table\eqref{tab:kappa_values}, show how the total mass is reduced to $M_s=4.975\cdot 10^{10}~\mathrm{kg}$ and the angle is $\alpha=20.5^\circ$. This configuration is referred to as deflective aluminium I.

\begin{comment}
\begin{table}[h!]
\centering
\resizebox{0.5\textwidth}{!}{%
\begin{tabular}{c c c c c c}
\hline

$t_{min}$ $[\mu m]$ & $\sigma$ [$kg/m$$^2$] &$\alpha [^\circ]$  & $\kappa$& $M_{\text{opt}}$ [kg] & $R_S$ [km] \\
\hline
 1 & 2.7$e-3$ & 20.5 & 0.175 &4.975e10 & 1430  \\
\hline
\end{tabular}
}
\caption{Parameters of a deflective aluminium sunshade with $\alpha = 20.5^\circ$.}
\label{tab:reference_case}
\end{table}
\end{comment}

The relevance of this formulation lies in the fact that the proposed framework remains parametric with respect to membrane thickness, which is a function of $r_s$ and $k$. As thinner membranes become technologically available, the same procedure can therefore be directly applied to update the design parameters and quantify the corresponding reduction in total launch mass.
%The resulting values for different prescribed thicknesses are reported in \ref{appendix:thickness}.

\subsection{Optimization of the sunshade equilibrium position}
\label{sec:optimization}
\begin{figure}[h]
    \centering
    \includegraphics[width=1\linewidth]{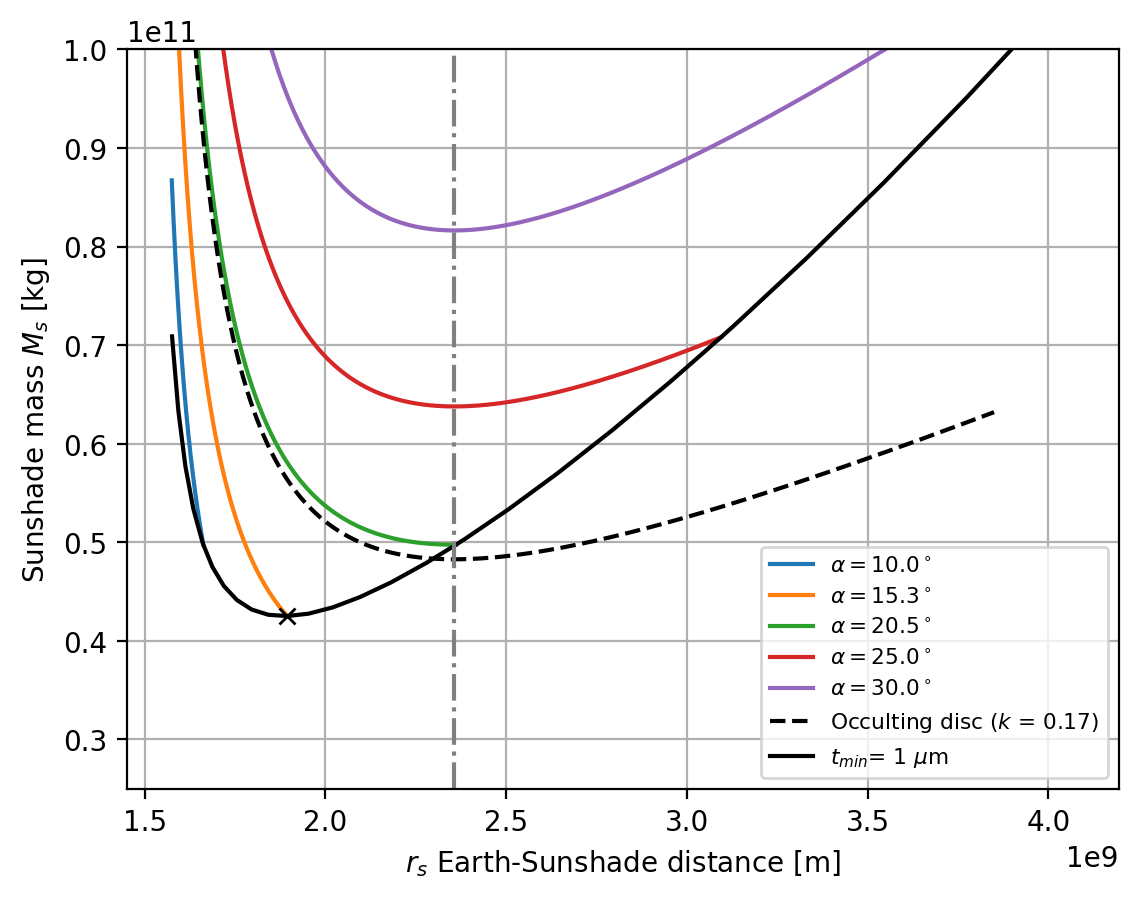}
    \caption{Sunshade mass $M_s$ as a function of the Earth-sunshade distance $r_s$ for different deflecting angles $\alpha$, assuming fixed aluminium optical properties $f_s=0.87$, $f_d=0.03$, and $f_a=0.10$. The dashed black curve represents the occulting-disc case with $\kappa=0.17$. The solid black curve denotes the admissible boundary imposed by the minimum membrane thickness constraint $t_{\min}=1~\mu\mathrm{m}$ for the deflective sunshade, while the black marker indicates the resulting minimum-mass configuration that can be achieved with $t_{\min}$ for a deflective sunshade.}
    \label{fig:fixed_thickness}
\end{figure}

As discussed in Sec.(\ref{preliminary_design}), for a fixed sunshade design the lowest mass is obtained at the optimal location $r_s^*$ if no other constraints are accounted for. In practice, and in particular in our case, manufacturability constraints may favour other locations that, while corresponding to higher masses for the same $\kappa$, allow the required material to be actually manufactured. 
in our case, for a given angle $\alpha$ of the deflective surface, varying the Earth-sunshade distance changes both the projected area required to produce the prescribed shadow at Earth and the associated membrane mass. 
In the unconstrained case, where the membrane thickness can be freely selected, the minimum mass solution is found at the optimal location $r_s^*$. 
However, when a lower bound on the membrane thickness is imposed, some configurations become infeasible because they would require an aluminium layer thinner than $t_{min}$ (see. Fig.\eqref{fig:fixed_thickness}).

We thus now consider a constraint on the minimum thickness of the aluminium membrane, $t_{\min}$, while keeping the optical properties fixed at $f_s=0.87$, $f_d=0.03$, and $f_a=0.10$, and assuming an aluminium density of $\rho=2700~\mathrm{kg/m^3}$. 
The corresponding resulting mass is shown in Fig.\eqref{fig:fixed_thickness} for $t_{\min}=1~\mu\mathrm{m}$, the reference value discussed above.
The Figure shows the sunshade mass for different values of $\alpha$, truncated where the required thickness falls below the imposed manufacturability limit. 
Under this constraint, we find that the minimum mass for an aluminium deflective sunshade improves to $\tilde{M_s} = 4.25 \cdot 10^{10}~\mathrm{kg}$ with $\tilde{\alpha} = 15.3^\circ$, $\tilde{r}_s = 1.89 \cdot 10^{9}~\mathrm{m}$ and consequently $\kappa=~0.128$ and $R_s=1150$ $\mathrm{km}$ (Eq.\ref{eq:radius}). This configuration is denoted as deflective aluminium II, and the corresponding values for $\tilde{\alpha}=15.3^\circ$ are reported in Table(\ref{tab:kappa_values}). For a fixed membrane thickness the areal density remains constant at $\sigma=\rho t=2.7\cdot10^{-3}~\mathrm{kg/m^{2}}$. Thus, the imposed thickness constraint is equivalent to an areal density constraint. Compared with the unconstrained optimum, the feasible optimum is shifted towards the Earth, i.e. closer to the $L_1$ point, and is associated with an angle $\tilde{\alpha}$ below the reference value of $20.5^\circ$. Moving the sunshade closer to Earth reduces the projected area required to achieve the same reduction in solar flux. Moreover, the resulting optimum mass for $\tilde{\alpha}=15.3^\circ$ is lower than that of the occulting disc.

\begin{figure}[h]
    \centering
\includegraphics[width=1\linewidth]{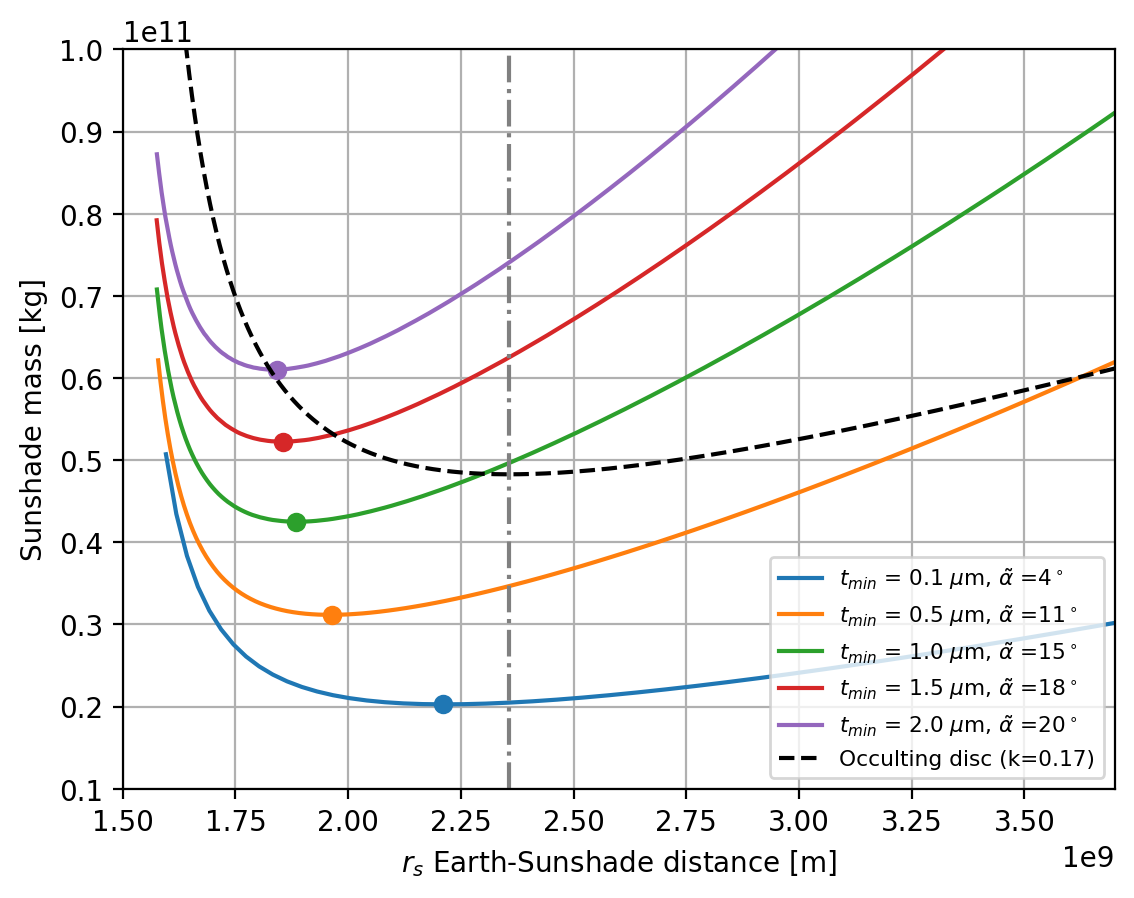}
  \caption{Sunshade mass as a function of the Earth-sunshade distance $r_s$ for different constrained membrane thicknesses. Markers denote the minimum-mass solution for each case, with the corresponding value of $\tilde{\alpha}$ indicated in the legend. The dashed black curve shows the reference occulting-disc configuration.}
    \label{fig:multiplethick}
\end{figure}

Fig.\eqref{fig:multiplethick} illustrates the effect of varying the imposed lower bound on the membrane thickness. When a smaller thickness is assigned, the minimum-mass solution progressively shifts closer to the classical optimum at $r_s^*$ while the mass tends to diverge when the thickness limit becomes bigger as the sunshade moves towards $L_1$. The plot also shows that the additional mass reduction obtained by changing the optimal sunshade position becomes less pronounced for smaller thickness constraints, as the solution approaches the classical optimum. By contrast, this benefit is more significant when larger membrane thicknesses are imposed. An analytical approximation for the location of the minimum-mass solution as a function of the thickness constraint and angle $\alpha$ is given in \eqref{Appendix:Reoptimisation}.

\subsection{Comparison of sunshade design configurations}
Table\eqref{tab:kappa_values} summarises the main sunshade concepts already mentioned in this work, reporting the corresponding angle $\alpha$ and material optical properties, which together determine the effective coefficient $\kappa$ through Eq.~\eqref{eq:kappa_eff}. The resulting sunshade mass $M_s$ is also provided, together with its ratio to the reference mass $\bar{M_s}^*$. The lower and upper bounds define the theoretical range of admissible configurations. As discussed in Sec.(\ref{sec:smallkappa}), the lower bound represents the minimum attainable value of $\kappa$, whereas the upper bound, $\kappa=1$, represents a perfect specular reflecting disc.  Intermediate values of $\kappa$ represent possible combinations of geometry and optical behaviour, including both existing and future sunshade concepts.

\begin{table}[h]
\centering
%\small
%\setlength{\tabcolsep}{0.1pt}
\begin{adjustbox}{width=\columnwidth}
\begin{tabular}{l c c c c c }
\hline
Configuration& $\alpha [^\circ]$& [$f_s,f_d,f_a$]   &$\kappa$ & $M_s$ [$\mathrm{kg}$] & $M_s$/$\overline M_s^*$\\
\hline
Lower-bound$^\dagger$  & 0.091& [$1,0,0$]  & $2.5e{-6}$ & 7.16e5 &  $2.5e{-6}$ \\
Upper-bound$^\dagger$  & 90& [$1,0,0$]  & $1$ & 2.84e11 &  $1$ \\
Reflective disc \cite{mcinnes} &90 &[$0.82,0,0.18$]  &  $0.91$ & 2.58e11 &   $0.91$\\
Occulting disc$^\dagger$ \cite{mcinnes} & 90 & [$0,0,1$] &  $0.17$ & 4.73e10   &   $0.17$  \\
Transparent disc$^\dagger$ \cite{transparent_occulters} & 90   & N/A &  $0.0019$  &  5.5e8  & $0.0019$ \\\hline
Deflective specular reflective $^\dagger$ & 20.5   &[$1,0,0$] & $0.12$ & 3.48e10 &   $0.12$   \\
Deflective aluminium I  & 20.5 & [$0.87,0.03,0.1$] & $0.175$ & 4.975e10& $0.175$\\
Deflective aluminium II & 15.3 & [$0.87,0.03,0.1$] & $0.128$ & 4.25e10& $0.15$ \\
%15.3& 1.89e9 & 1 & 2.7$e-3$  & 0.128 & 4.25e10 & 0.15  & 1150 & 4204 
\hline
\end{tabular}
\end{adjustbox}
\caption{Characteristic parameters of the different sunshade configurations considered in this study.
The configurations marked with ($^\dagger$) are theoretically admissible but not considered realistic for the assumed material properties.}
\label{tab:kappa_values}
\end{table}

\begin{figure}[h]
    \centering
    \includegraphics[width=1\linewidth]{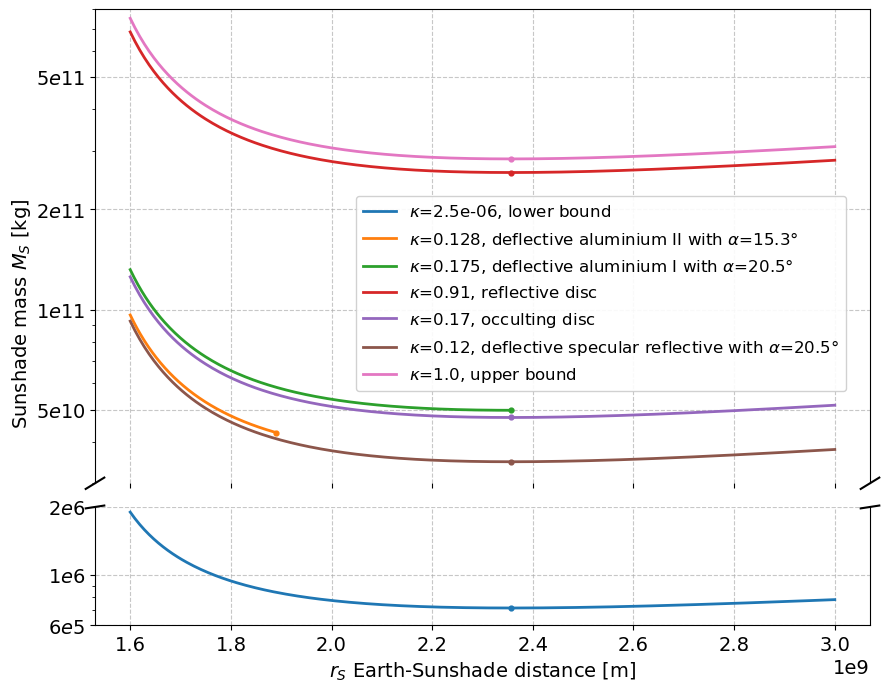}
    \caption{Comparison of sunshade configurations for different values of $\kappa$. The sunshade mass $M_s$ is shown as a function of the sunshade position $r_s$, with markers denoting the minimum achievable mass for each configuration.}
    \label{fig:comparison}
\end{figure}

This trend is also illustrated in Fig.\eqref{fig:comparison}, where the sunshade mass is plotted as a function of the sunshade distance from Earth $r_s$ for different values of $\kappa$. The two limiting cases lead to optimal mass values that differ by more than five orders of magnitude, highlighting the strong sensitivity of the required sunshade mass to $\kappa$.  Moreover, an ideal specular deflective sunshade performs better than an aluminium deflective sunshade with the same geometry, due to the non-ideal optical properties of aluminium. The deflective aluminium I configuration, with $\alpha=20.5^\circ$, achieves a performance comparable to that of the occulting disc, whereas the deflective aluminium II with $\tilde{\alpha}=15.3^\circ$ outperforms both cases. The curves corresponding to deflective aluminium I and deflective aluminium II are plotted only over the range of $r_s$ values for which the configurations are feasible under the imposed minimum-thickness constraint ($t_{\min}=1~\mu\mathrm{m}$). Outside these ranges, the same configurations would require a membrane thickness below this limit and are therefore excluded from the feasible design space.

\subsection{\label{sec:finalcone}Conical configuration as a representative deflective sunshade}

The results presented in Sec.(\ref{sec:optimization}) indicate that, within the assumed technological limits, the lowest mass for an aluminium deflective sunshade is obtained for $\tilde{\alpha}=15.3^\circ$. This concept can be implemented through different geometrical layouts; among the possible geometrical implementations, a hollow conical surface with no base membrane is considered here as a representative configuration. This geometry is consistent with the general deflective sunshade formulation because the required angle $\alpha$ coincides with the cone half-opening angle. Thus, the optimal value of $\alpha$ is imposed by the overall geometry itself, rather than more complex structural arrangements. The conical configuration is therefore adopted as a reference case owing to its geometrical simplicity, which provides a convenient basis for the following discussion of deployability and foldability, as will be discussed in Sec.(\ref{sec:coneorigami}). The cone axis is aligned with the Sun--Earth line and the base radius coincides with the required sunshade radius of $R_s=1150$ $\mathrm{km}$, while the cone height follows directly from the cone geometry with $h_{cone}=4204$ $\mathrm{km}$. 

\begin{comment}
\begin{table}[h!]
\centering
\begin{adjustbox}{width=\columnwidth}
%\resizebox{0.49\textwidth}{!}{%
\begin{tabular}{c c c c c c c c c c}
\hline

$\alpha [^\circ]$  & $r_s$ [m] & $t_{min}$ [$\mu m$] & $\sigma$ [$kg/m$$^2$] & $\kappa$& $M_s$ [kg] &  $M_s$/$\overline M_s^*$ & $R_S$ [km] & $h_{cone}$ [km] \\
\hline

20.5 &  2.36e9   & 1  & 2.7$e-3$     & 0.175  & 4.975e10 & 0.175 & 1430   &  3825 \\  
  \hline
15.3& 1.89e9 & 1 & 2.7$e-3$  & 0.128 & 4.25e10 & 0.15  & 1150 & 4204 \\

\hline
\end{tabular}
\end{adjustbox}
\caption{Parameters of .....}
\label{tab:cone}
\end{table}
\end{comment}

\section{\label{sec:coneorigami}Origami structures for conical shapes}

Besides manufacturing, deploying a structure of the size proposed in Sec.(\ref{sec:finalcone}) remains a major challenge. Instead of considering a single large cone, one may instead consider a swarm of smaller units with the same total obscured area and therefore the same overall effect. Even with a federated sunshade, however, the sheer amount of mass that must be launched calls for efficient methods to package and deploy these cones. In this context, origami-inspired structures offer compact folding for storage and transport, while enabling deployment to large and complex shapes with high stiffness-to-weight ratios. We therefore consider origami-based folding of these cones and analyse the sizing and deployment of the resulting structures. A main bottleneck in the sizing process is the fairing volume. At the time of writing, the largest claimed payload capacity is that of Starship \cite{Starship}, which is reported to offer a payload capacity of up to 150 tonnes with a fairing approximately $9~\mathrm{m}$ in diameter and $18~\mathrm{m}$ in height.

Our origami construction of choice folds the cone completely flat in its initial state. Together with the very thin aluminium shell of the cone itself, the main size constraint is therefore the fairing radius of $4.5~\mathrm{m}$.
We follow the construction by Sharma and Upadhyay \cite{sharma2021folding} for a conical origami pattern based on the Miura--Ori four-fold-line origami. Other origami patterns for conical shapes exist, such as Kresling origami \cite{lu2022conical}, but these typically sacrifice rigidity \cite{kidambi2020dynamics}, the ability of an origami structure to be folded entirely from rigid sheets without buckling or bending, or developability \cite{misseroni2024origami}, meaning that the structure can be manufactured from a single sheet.
On the other hand, the Miura-Ori pattern has been widely used for space structures \cite{ravichand2024origami} and offers excellent properties, such as a single degree of freedom, rigidity, developability and flat foldability \cite{misseroni2024origami} \cite{fang2017dynamics}. Another excellent property of this specific pattern is that the unfolded radius exceeds the folded radius of the cone. This means that while folded, the cone has to satisfy $r < 4.5~\mathrm{m}$ but can be larger unfolded. We will use this property together with choosing beneficiary parameters to unfold the cone into slightly bigger size while still fitting inside the Starship fairing.

\subsection{Construction of a cone origami}

The origami pattern by Sharma et al. discretizes the conical shape by considering an $m$-gonal Pyramid and cutting it vertically into segments. The pattern does not account for the pointed apex, so the final origami cone will always have an open top, but one can increase the number of segments to decrease the size of this hole to an arbitrary size. The segments are mirrored and resized versions of one segment, and each segment consists of the same number of unit cells.

\begin{figure}[htbp] 
        %\centering
        \includegraphics[width=\linewidth]{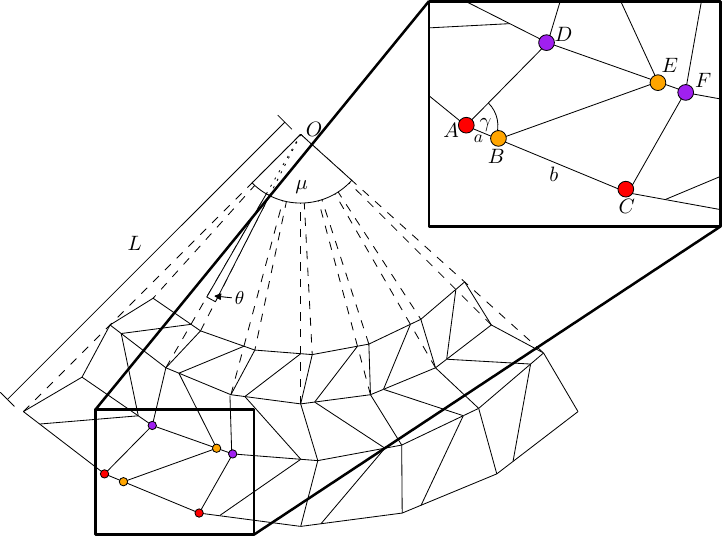}
        \hfill
        \caption{Net of the cone origami with $3$ segments and $6$ units per segment. Visualised is the shift angle $\theta$, slant length $L$ and arc angle $\mu$. The arc angle $\mu$ for a net of a cone with angle $\alpha$ is $\mu = 2 \pi \sin{(\alpha)}$. \\ In the top right corner, a unit cell is displayed, the marked points show symmetries between the different vertices of the unit cells. Visualised is also the folding angle $\gamma$.}
        \label{fig:cone_net_cell}

    % red and purple points are the different vertices of the boundary $m$-gons of each segment and orange points are placed such that the connecting edges have a ratio of $lr = \frac{a}{b}$

    %\begin{subfigure}{\columnwidth}
        %\centering
    %    \includegraphics[width=\linewidth]{Figures/cone_cell.pdf}
    %    \hfill
    %    \caption{A single unit cell of the cone origami. The marked points show symmetries between the different vertices of the unit cells; red points and purple points are the different vertices of the boundary $m$-gons of each segment and orange points are placed such that the connecting edges have a ratio of $lr = \frac{a}{b}$. Visualized is also the folding angle $\gamma$.}
    %    \label{fig:cone_cell}
    %\end{subfigure}
\end{figure}

In addition to the cone parameters, height $h$ and half-angle $\alpha$, an origami segment is additionally characterised by four more parameters, units per segment $m$, fold angle $\gamma$, length ratio $lr$ and shift angle $\theta$. We will also fix a number of segments $n$.
More specifically, consider a cone with already fixed height $h$ and half-angle $\alpha$ and consider its net (Fig.\eqref{fig:cone_net_cell}) for which we get the slant length $L$ and arc angle $\mu$ by
$$
    \mu = 2 \pi \sin \alpha \quad \text{and} \quad L = \frac{h}{\cos \alpha}.
$$

As shown in Fig.\eqref{fig:cone_net_cell}, it is possible to cut the net into triangular sectors depending on the number of units per segment wanted. We therefore have $m + 1$ rays from the apex to each end point of the triangular sections. To find the next segment, construct independently the same $m+1$ rays from the origin but shifted by the shift angle $\theta$ as well as for each of the $m+1$ endpoints, a ray with a fold angle of $\gamma$ from the line segment between it and the next endpoint upward. Pairing up each of the two sets of $m+1$ new lines leads to $m+1$ intersection points that are the basis for the next segment. 
This construction can be repeated by alternating the shift direction to get more segments in a zig-zag pattern. The origami net should now look like a collection of quadrilaterals. Each quadrilateral is considered a unit cell. 
The final step in the construction is adding another fold to each unit cell: Consider an arbitrary unit cell with end points $A,C,D$ and $F$ with $A,C$ belonging to one segment and $D,F$ to the next. Note that the line segment $AD$ is at an angle of $\gamma$ from $AC$ as constructed. Find point $B$ between $A$ and $C$ such that the length of $AB =: a$ and $BC =: b$ have a ratio of $\frac{a}{b} = lr$. Similarily but mirrored, find point $E$ on $DF$ such that $DE =: c$ and $EF =: d$ have ratio $\frac{d}{c} = lr$. One again alternates this construction in a zig-zag pattern so that the ratio inverts and the line segment is therefore connecting to the previous endpoint. 

This construction is visualised in Fig.\eqref{fig:cone_net_cell}.

\begin{figure}[htb]
    \centering
    \includegraphics[width=1\linewidth]{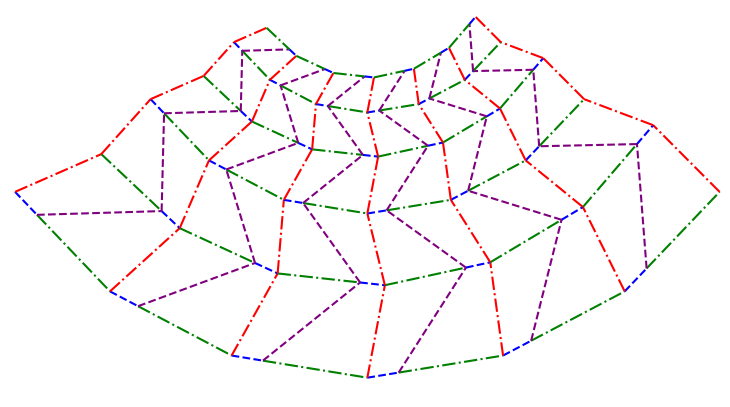}
    \hfill
    \caption{A finished origami cone net. Dashed lines are valley folds and dot-dashed lines are mountain folds. Colors group together different types of line segments that are pairwise similar. One can then assign to each type a fold type: \\
    (a) Red line segments denote the $m+1$ lines from each segment base to the apex or equivalently from each $A$ point to its respective $D$ point. This type will always be a mountain fold. \\
    (b) Purple line segments denote the $m+1$ lines going from each $B$ to its respective $E$ point and will always be valley folds. \\
    (c) Blue folds denote the shorter of the two line segments connecting the $B$ points to either $A$ or $D$. During the construction the ratio of the length of the blue line segment with respect to the green one equals the parameter $lr$. These folds will be valleys.\\
    (d) Green folds denote the longer of the two line segments connecting the $B$ points to either $A$ or $D$. This is the counterpart to the blue fold lines and will therefore be mountain folds.}
    \label{fig:cone_net_origami}
\end{figure}

Finally, we follow the Miura-Ori pattern and assign folds in such a way that each vertex has either three mountain folds and one valley fold or vice-versa.
The final origami cone net can be seen in Fig.\eqref{fig:cone_net_origami}.

\subsection{\label{sec:origamiparameterselection}Selection of Origami parameters}

Whilst a full scale structural simulation model is out of the scope of this work, we can still choose the origami parameters in a suitable manner to aid in deployment and integrity. In an ideal origami, no crease or facet experiences deformation during the folding and unfolding. The 2D-Miura-Ori pattern for a flat sheet posses this property \cite{misseroni2024origami} but unfortunately the proposed Miura-Ori cone does not. For origami patterns, the geometrical Cauchy strain, defined as the ratio between the deformation of a crease during unfolding and its starting length, has shown to be a good estimator of the real engineering strain on the physical origami \cite{ghassaei2018fast}. As the Cauchy strain measures deformation, we want this to be minimised in order for our material to experience the least deformation.
Interested readers are referred to studies by Sharma et al.~\cite{sharma2021folding} in which the deformation during unfolded is simulated with a pin-jointed truss model of the origami structure.

Besides the work by Sharma et al., we will supplement the studies on parameters with our own simulations using a geometric origami solver \cite{ghassaei2018fast}. The simulation tool by Ghassaei et al. considers the origami as a rigid-body linkage where each crease is folded simultaneously by the same relative folding angle, meaning if the origami is $40\%$ folded, then the current folding angle of every crease is $40\%$ of the final folding angle, assuming they all start at $0^\circ$. In an unidealised origami, there might be no configuration of the rigid-body with this property so the origami solver considers small displacements of each node to find the least displacement solution  that has the correct folding angles, while the total displacement on the pattern, defined by the sum of displacements of each crease, is minimised. For this least displacement solution, we can then calculate the geometrical Cauchy strain of the creases and nodes by calculating their ratio between deformation and starting length.
For a fixed set of parameters, the geometrical Cauchy strain of that origami is the maximum geometrical Cauchy strain of any crease during any unfolding step. For more details, see the work of Ghassaei et al. \cite{ghassaei2018fast}.

Starting with fixed folded radius $R$ and cone half angle $\alpha$, we start from the optimal configuration suggested by Sharma et al. \cite{sharma2021folding} and discretely vary the parameters such that geometrical Cauchy strain during unfolding is minimised.

The geometrical Cauchy strain of a crease in the origami pattern with initial length $l_0$ and deformed length $l_t$ at normalised timestep $t \in [0,1]$ is 
\[
    \frac{l_t-l_0}{l_0}
\]
and measures the relative displacement of this crease. At a node with $N$ creases $l_1,\dots,l_N$ the strain is calculated as
\[
    \frac{1}{N} \sum_{i=1}^N \frac{\Delta l_i}{l_i}
\]
and measures the average of the strains of its creases. Finally, on each face, an interpolation of the strains of the delimiting beams and nodes is displayed. For more details on this, see the work by Ghassaei et al. \cite{ghassaei2018fast}.

\begin{figure}[htbp]
    \centering 

    % ---------- Row 1 ----------
    \begin{subfigure}{0.48\columnwidth}
        \centering
        \includegraphics[width=\linewidth]{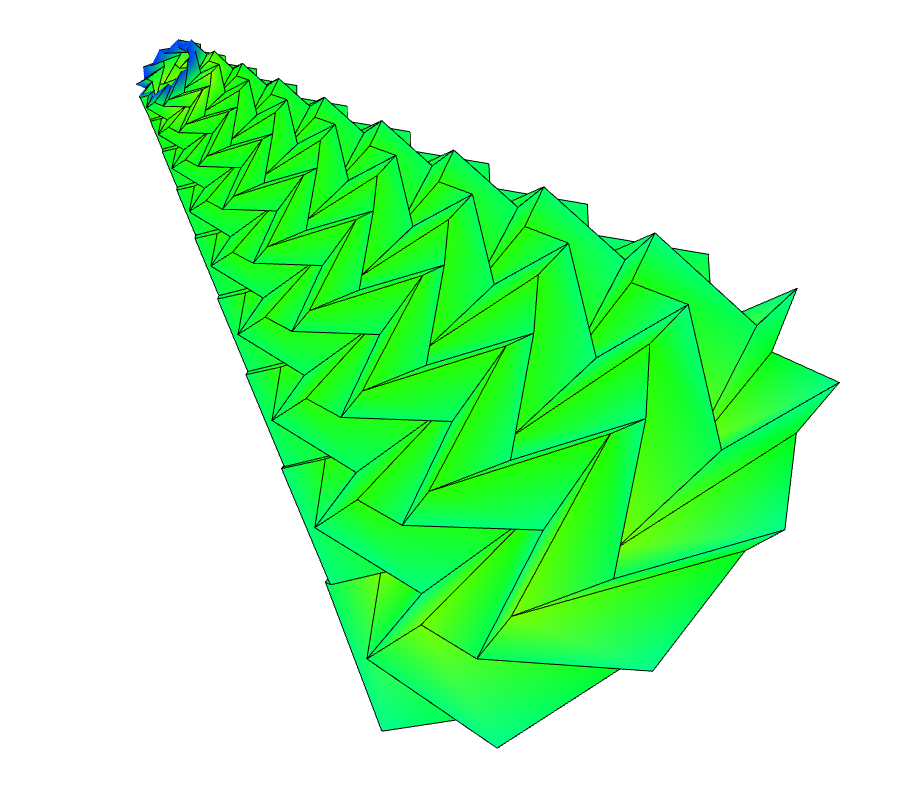}
        \caption{}
        \label{fig:cone_strain_base}
    \end{subfigure}
    \hfill
    \begin{subfigure}{0.48\columnwidth}
        \centering
        \includegraphics[width=\linewidth]{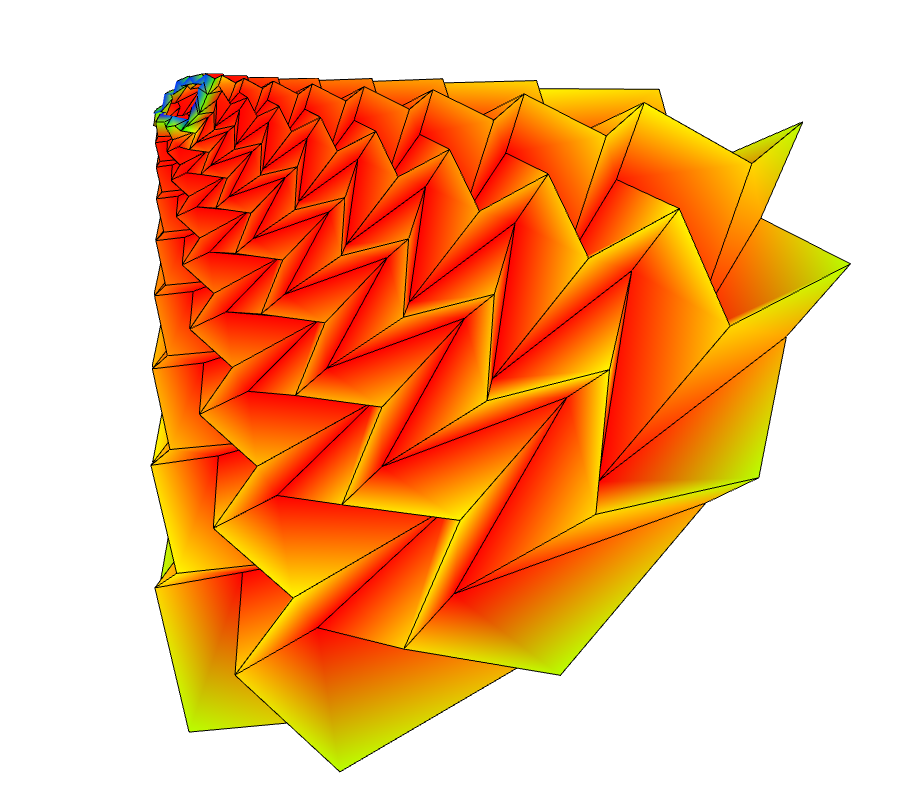}
        \caption{}
        \label{fig:b}
    \end{subfigure}

    \vspace{0.2cm}

    % ---------- Row 2 ----------
    \begin{subfigure}{0.48\columnwidth}
        \centering
        \includegraphics[width=\linewidth]{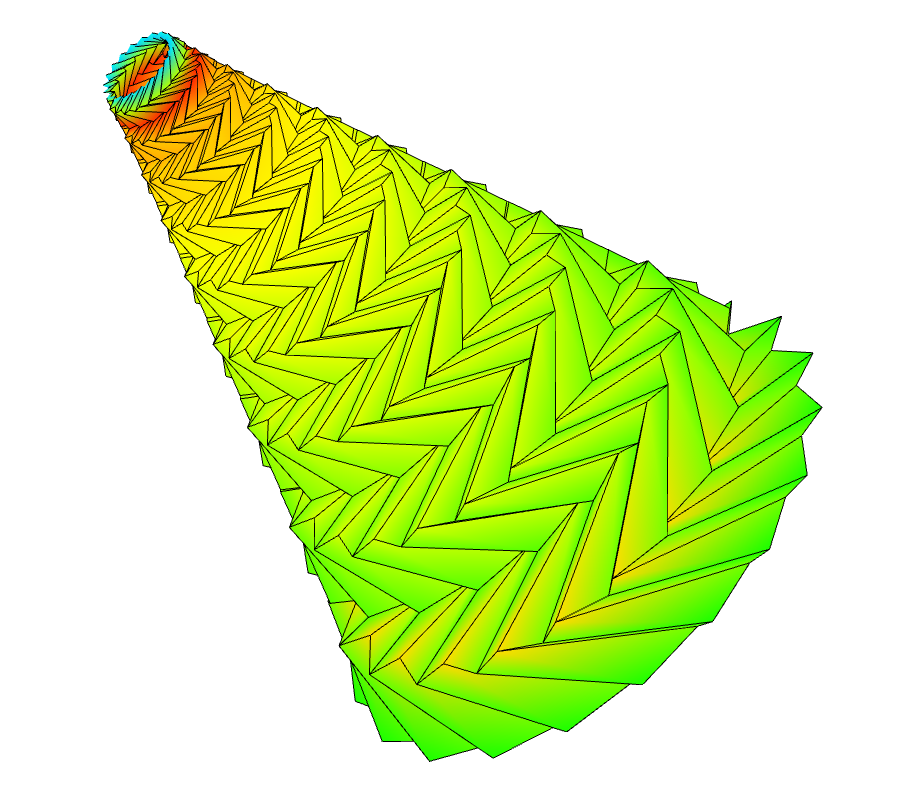}
        \caption{}
        \label{fig:cone_strain_units}
    \end{subfigure}
    \hfill
    \begin{subfigure}{0.48\columnwidth}
        \centering
        \includegraphics[width=\linewidth]{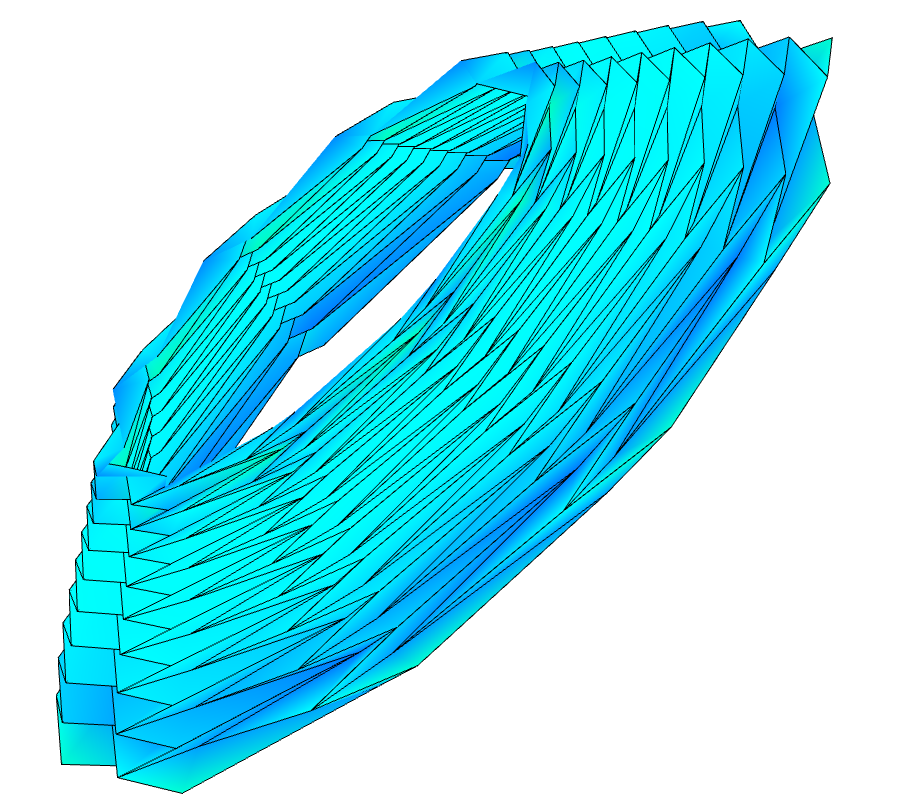}
        \caption{}
        \label{fig:cone_strain_fold_angle}
    \end{subfigure}

    \vspace{0.2cm}

    % ---------- Row 3 ----------
    \begin{subfigure}{0.48\columnwidth}
        \centering
        \includegraphics[width=\linewidth]{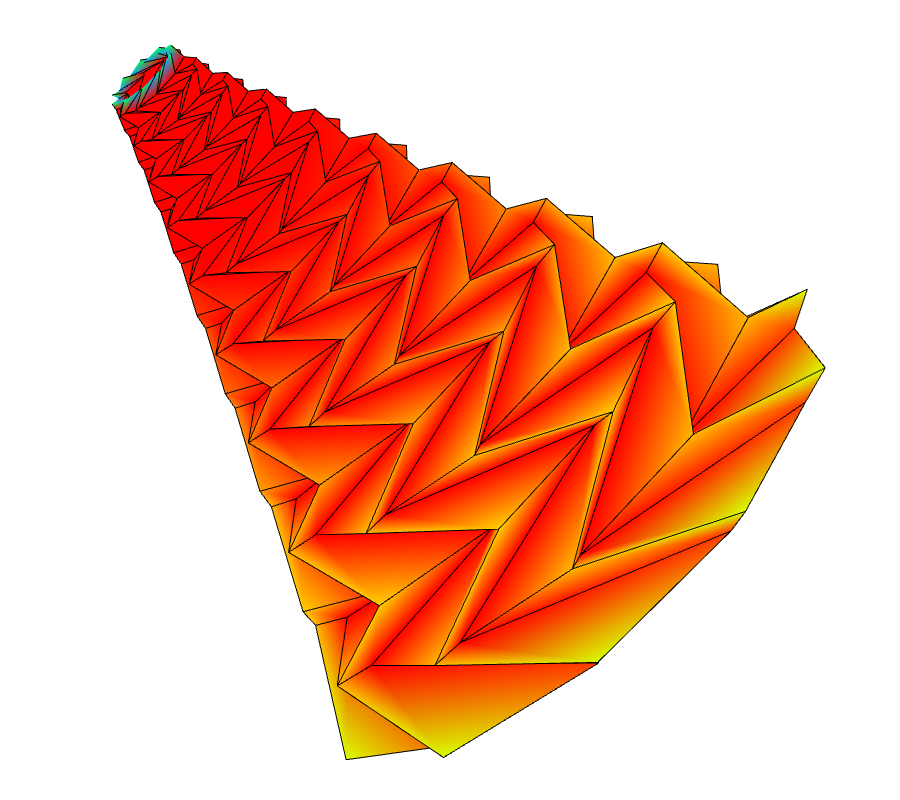}
        \caption{}
        \label{fig:cone_strain_length_ratio}
    \end{subfigure}
    \hfill
    \begin{subfigure}{0.48\columnwidth}
        \centering
        \includegraphics[width=\linewidth]{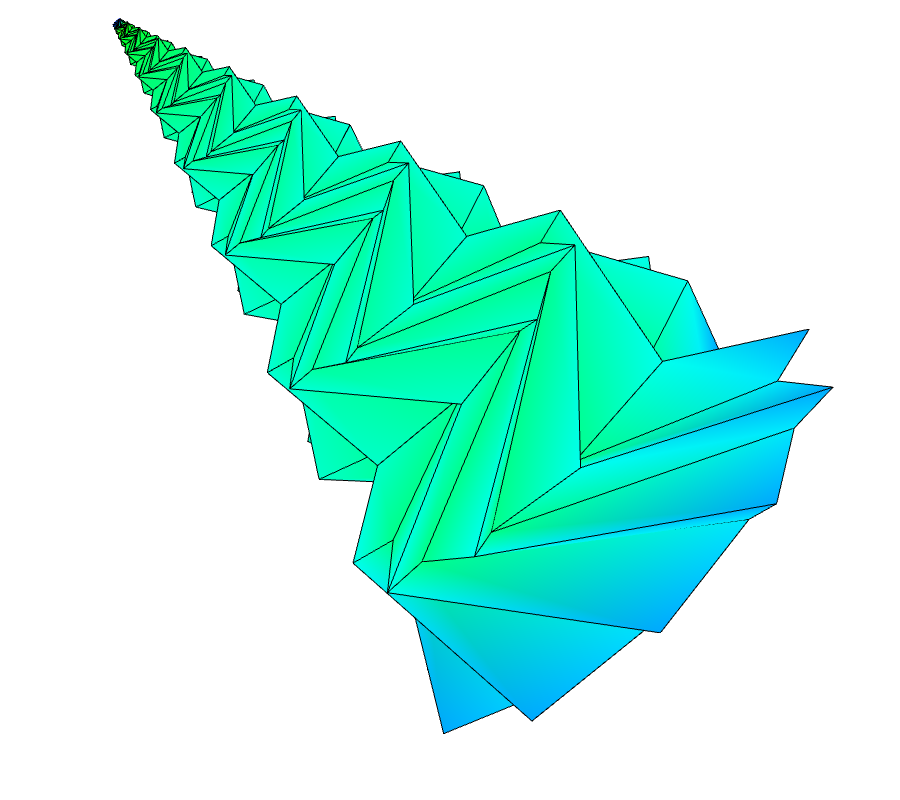}
        \caption{}
        \label{fig:cone_strain_shift_angle}
    \end{subfigure}

    \vspace{0.1cm}

    % ---------- Long image ----------
    \begin{subfigure}{0.98\columnwidth}
        \centering
        \includegraphics[width=\linewidth]{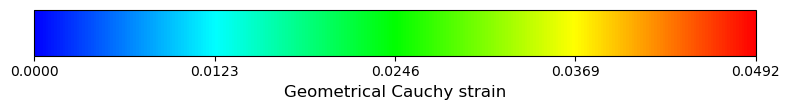}
        \label{fig:cauchy_strain_scale}
    \end{subfigure}

    % ---------- Global caption ----------
    \caption{
    Geometrical cauchy strain during unfolding on cones with different parameters. The strain is calculated as the ratio between the deformation of a crease to its starting length. The strain is computed for each crease and interpolated on the faces. \\
    (a) The base case with 18 segments, 10 units per segment, a cone half angle of 15.3°, a fold angle of 63°, shift angle of 3° and length ratio of 0.33. \\
    (b) Cone half angle of 30° instead of 15.3°. \\
    (c) 20 units per segment instead of 10. \\
    (d) Fold angle of 30° instead of 63°. \\
    (e) Length ratio of 0.2 instead of 0.33. \\
    (f) Shift angle of 6° instead of 3°.
    }
    
    \label{fig:cauchy_strain}
\end{figure}

The strain corresponding to different parameter sets is visualised in Fig.\eqref{fig:cauchy_strain}. Our findings overlap with those of Sharma et al. \cite{sharma2021folding} in all parameters with the exception of length ratio in which we found an optimal value at $0.33$ instead of the proposed $0.2$. We quickly summarize the effects of the different origami parameters on the folded and unfolded radius and strain. A more complete study on the influence of these parameters on the displacement and radius, including an extensive set of plots, has been done by Sharma et al. and we refer interested readers to them \cite{sharma2021folding}.

Increasing the number of units per segment $m$ decreases the folded radius $R$ but might cause more units to overlap in the flat-folded state, increasing the thickness of the folded origami.

As the folding angle $\gamma$ increases, the folded radius $R$ decreases. This effect can be observed until $\gamma$ reaches an upper bound of about $63^\circ$ degrees. Starting at $\gamma \geq 72^\circ$ degrees, the flat folded origami self-intersects and is no longer physically feasible.

The length ratio $lr$ decreases the folded radius $R$ as it decreases but starts to rapidly increase the cauchy strain of the bottom segments when below $0.33$.

Variations in the shift angle $\theta$ do not influence the final folded radius $R$, due to only influencing the folding structure of the second segment and onward. However a smaller shift angle leads to rapidly increasing strain during unfolding and more overlapping segments in the folded state, while a larger shift angle leads to increased torsion in the units. Above a threshold of about $8^\circ$ degrees, this leads to the origami not being flat foldable due to the units self-intersecting during folding.

Considering this, we chose $m = 10$, $\gamma = 63^\circ$ degrees, $lr = 0.33$ and $\theta = 3^\circ$ degrees. With these parameters, the geometrical Cauchy strain during all unfolding angles was less than $2.1\%$, meaning we can expect the material to deform less than $2.1\%$ during unfolding, and the folded radius was about $82.4\%$ of the unfolded radius. Additionally, we chose the number of segments such that the open top in the unfolded state has a radius of less than $10\%$ of the radius of the unfolded bottom $m$-gon. For our choice of parameters, this was achieved after having at least $19$ segments.

\section{\label{sec:system}Preliminary system analysis}
\noindent 

\noindent The dimensions of the final design, reported in Sec.(\ref{sec:finalcone}), motivate the adoption of a constellation of smaller sunshades.
The origami-inspired folding pattern presented in Sec.(\ref{sec:coneorigami}) is an example of how one could stow individual cones within the fairing of the selected launch vehicle. 

The next-generation Starship vehicle \cite{Starship} by SpaceX is designed to provide a payload capacity up to 150 tonnes for low Earth orbit. Starship is expected to have a payload fairing approximately of $9~\mathrm{m}$ in diameter and $18~\mathrm{m}$ in height, which would provide one of the largest usable payload volumes among launch vehicles currently operational or under development \cite{fairingstarship}. It is therefore adopted as the reference launcher in this analysis. The folding constraint is imposed by the maximum allowable stowed radius of $4.5\,\mathrm{m}$. However, as shown in Sec.\eqref{sec:coneorigami}, the origami folding strategy leads to a single cone having a larger unfolded radius of $R_s = 5.46\,\mathrm{m}$ while still having a folded radius of $4.5\,\mathrm{m}$. This corresponds to a cone height of approximately 19.95 \,m and, consequently, to a required total number of cones of approximately $4.4\cdot10^{10}$, each of a $M_{i}=0.96~\mathrm{kg}$. However, this estimate only includes the aluminium membrane and does not account for the structural frame or any additional subsystems that may be necessary. For the fixed total system mass, introducing a non-negligible frame mass per cone would therefore reduce the number of units that could be deployed, as part of the available mass budget would be allocated to the supporting structure.
Considering a payload capacity of 150 tons per launch, the required number of launches would be approximately $ 2.83 \cdot 10^5$. The current rate is approximately one launch every three days. For instance, Cape Canaveral’s launch schedule has expanded considerably over the past four years, rising from a historical average of about 15 orbital missions per year to more than 50. The turnaround time between launches is now frequently less than a week, and on some occasions, two launches have occurred on the same day\cite{launchcapacity}. If this upward trend continues, launch rates can be expected to rise further in the coming years. 

\begin{figure}[!ht]
    \centering
    \includegraphics[width=1\linewidth]{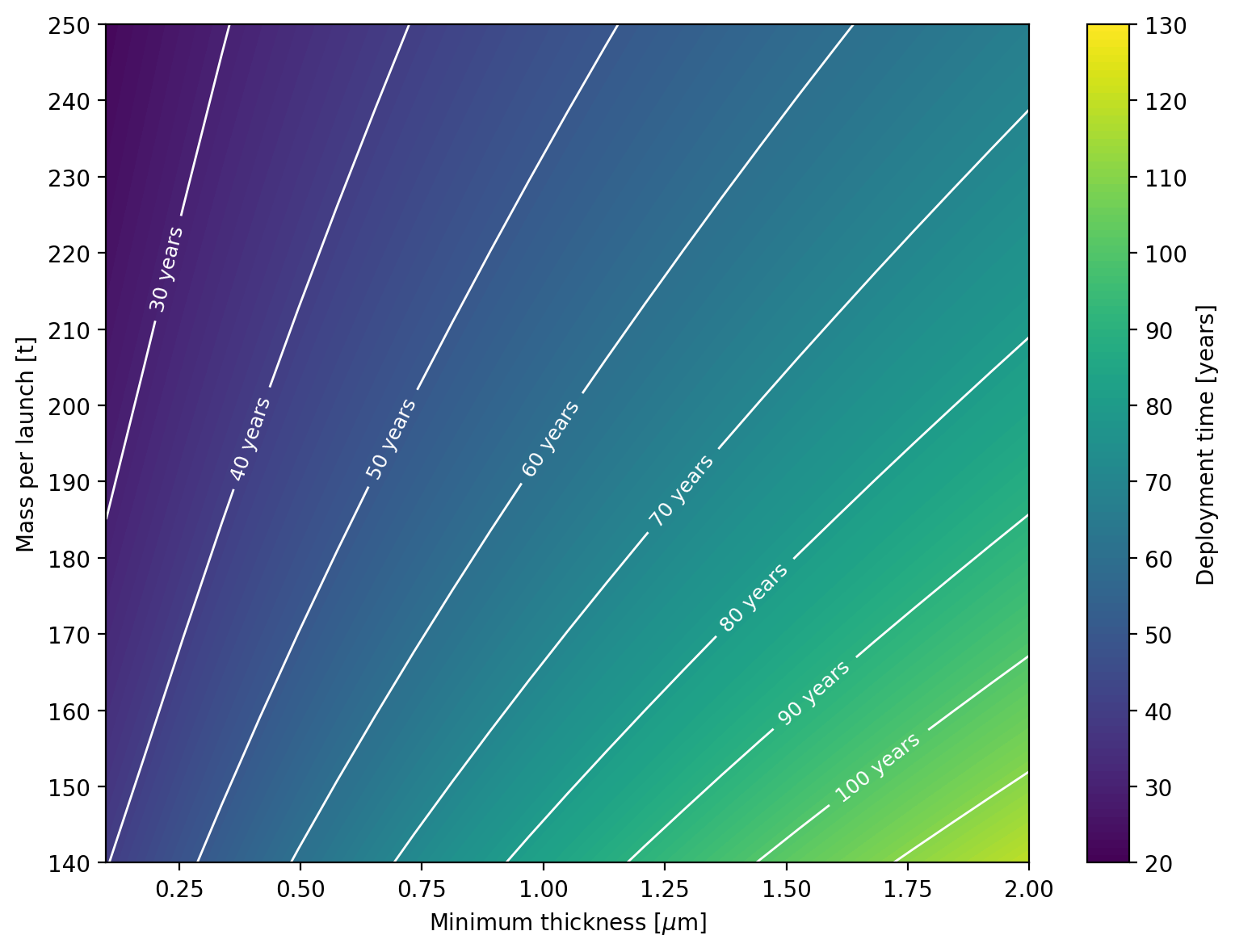}
    \caption{Deployment time as a function of the constrained thickness $t_{min}$ for different launcher payload capacities, assuming 10 launches per day. The aluminium sunshade position is equal to the corresponding $\bar{r_s}$ calculated following the approach explained in Sec \ref{sec:optimization}.}
    \label{fig:launches_thickness}
\end{figure}

Fig.(\ref{fig:launches_thickness}) shows the total deployment time as a function of the minimum thickness constraint and of different launcher payload, assuming a launch cadence of 10 launches per day. This launch rate is not currently achievable, and although the launch frequency is increasing over time, it is not possible to predict when such a cadence could realistically be reached. Therefore, this value should be interpreted as a reference parameter: if the launch rate were reduced by a factor of two, the deployment time would simply double. 

\begin{comment}
\begin{figure}[!ht]
    \centering
    \includegraphics[width=1\linewidth]{Figures/numberlaunches.png}
    \caption{Deployment time as a function of cone angle $\alpha$ for different launcher payload capacities, assuming 10 launches per day. In this case the aluminium sunshade position is fixed at $r_s^*$.}
    \label{fig:numberlaunches}
\end{figure}
\end{comment}

As an example, assuming the cone angle  $\tilde{\alpha}=15.3^\circ$ and a payload capacity of 150 tonnes per launch, the resulting deployment time is almost 80 years. However, future technological improvements could significantly reduce this value. For example, the ability to manufacture thinner aluminium foils would reduce the total mass and, consequently, the number of launches.
At the same time, launch vehicles are expected to continue improving in terms of payload capacity. The use of higher capacity launchers would further reduce the total number of launches. For instance, assuming a payload capacity of 250 tonnes per launch and a minimum thickness of $0.25~\mu \mathrm{m}$, the deployment time would decrease to less than 30 years. Therefore, while the project is not possible in the short term with current launch capabilities, the design itself appears technically feasible. 
On the other hand, several major challenges remain to be addressed. The control of the conical structure requires further investigation, particularly with respect to attitude stability and station-keeping. In addition, the deployment and coordination of a constellation composed of a very large number of individual structures would introduce substantial operational complexity. Further work is also needed to assess the manufacturing, launch, assembly, and deployment logistics associated with producing and operating such a large total membrane area.

%%%%%%%%%%%%%%%%%%%%%%%%%%%%%%%%%%%%%%%%%%%%%%%%%%%%%%%%%%%%%%%%%%%%%%%%%%%%%
% \clearpage
\section{\label{sec:conclusions}Conclusion}
\noindent In this work, we have developed a general framework for the analysis and preliminary design of deflective space-based sunshades, namely sunshades that reduce the required launch mass by shaping the solar-radiation-pressure response through macroscopic geometry rather than by relying solely on advanced materials. 
Starting from a standard mass-balancing formulation, we reduced the whole design to engineering a favourable effective momentum-transfer coefficient and recovered as limiting cases several previously studied sunshade concepts. 
Within this framework, we examined a deflective architecture based on inclined aluminium reflective surfaces. Under ideal specular assumptions, the minimum mass is achieved for small inclination angles, confirming the strong advantage of redirecting rather than blocking sunlight. 
When realistic aluminium optical properties and a finite manufacturable thickness are imposed as constraints, the attainable performance is significantly reduced, but the concept remains lighter than a classical occulting disc. 
Assuming a minimum aluminium thickness of $1~\mu\mathrm{m}$, the optimal configuration results in a deflective angle $\tilde{\alpha} = 15.3^\circ$ and a distance $\tilde{r}_s = 1.89 \cdot 10^6~\mathrm{km}$ along the Earth-Sun line, leading to a total mass of $M_s = 4.25 \cdot 10^{10}~\mathrm{kg}$.
A conical geometry was then studied as a representative implementation of this deflective concept. This choice provides a simple axisymmetric realization of the general design principle and yields a practical reference case for both system sizing and deployability analysis. 
Because the required scale is far beyond what a monolithic structure could accommodate, we considered a constellation of smaller conical units with equivalent total area and mass. To address packaging and deployment constraints, we proposed an origami-inspired flat-folding strategy based on a generalised Miura--Ori pattern adapted to the conical geometry. 
Overall, our results suggest that deflective sunshades may provide a promising avenue for reducing the mass penalty associated with space-based solar radiation management. 
Although the specific architecture examined here is not yet likely to constitute a near-term engineering solution, it establishes a reference design that can be manufactured and against which future innovations in sunshade concepts can be assessed. 
More broadly, this work introduces deflective sunshade geometry as an additional design option, develops a unified framework that recovers and generalizes previous deflective-sunshade formulations, identifies the principal geometric–optical trade-offs, and outlines a basis for comparative studies of future sunshade architectures.

%%%%%%%%%%%%%%%%%%%%%%%%%%%%%%%%%%%%%%%%%%%%%%%%%%%%%%%%%%%%%%%%%%%%%%%%%%%%%
\section*{Acknowledgements}
 The authors gratefully acknowledge Quirien Wijnands for the initial discussions that rekindled our interest in geo-engineering and ultimately led to the inception of this project.

%%       \cite{<label>} ==> [1]

%%
%% Example citation, See \cite{lamport94}.


\bibliographystyle{elsarticle-num} 
\bibliography{lib.bib}

@misc{ParisAgreement,
  author = {{United Nations Framework Convention on Climate Change}},
  title  = {The Paris Agreement},
  year   = {2026},
  note   = {\url{https://unfccc.int/process-and-meetings/the-paris-agreement} Accessed: 2026-06-12}
}

@article{geo,
  author  = {Schneider, Stephen H.},
  title   = {Earth systems engineering and management},
  journal = {Nature},
  year    = {2001},
  volume  = {409},
  pages   = {417--421},
  doi     = {10.1038/35053203}
}

@article{firstsunshade,
  author  = {Early, James T.},
  title   = {Space-based solar shield to offset greenhouse effect},
  journal = {Journal of the British Interplanetary Society},
  year    = {1989},
  volume  = {42},
  pages   = {567--569}
}

@misc{NASA_carbon_dioxide,
  author = {{NASA Science}},
  title  = {Carbon Dioxide -- Earth Indicator},
  year   = {2026},
  note   = {\url{https://science.nasa.gov/earth/explore/earth-indicators/carbon-dioxide/} Accessed: 2026-06-12}
}

@misc{NASA_global_temperature,
  author = {{NASA Science}},
  title  = {Global Temperature},
  year   = {2026},
  note   = {\url{https://science.nasa.gov/earth/explore/earth-indicators/global-temperature/} Accessed: 2026-06-12}
}

@article{sharma2021folding,
  title={Folding pattern design and deformation behavior of origami based conical structures},
  author={Sharma, Hemant and Upadhyay, Sanjay H},
  journal={Advances in Space Research},
  volume={67},
  number={7},
  pages={2058-2076},
  year={2021},
  publisher={Elsevier}
}

@article{mcinnes,
author = {McInnes, C.},
year = {2010},
month = {03},
pages = {571-580},
title = {Space-based geoengineering: Challenges and requirements},
volume = {224},
journal = {Journal of Mechanical Engineering Science Journal of Mechanical Engineering Science},
}

@article{struck,
  author  = {Struck, C.},
  title   = {The feasibility of shading the greenhouse with dust clouds at the stable lunar Lagrange points},
  journal = {Journal of the British Interplanetary Society},
  year    = {2007},
  volume  = {60},
  pages   = {82--89}
}

@article{angel,
author = {Angel, Roger},
year = {2006},
month = {12},
pages = {17184-9},
title = {Feasibility of cooling the Earth with a cloud of small spcecraft near the inner Lagrange point (L1)},
volume = {103},
journal = {Proceedings of the National Academy of Sciences of the United States of America},
}

@article{tether,
author = {Szapudi, István},
year = {2023},
month = {07},
pages = {e2307434120},
title = {Solar radiation management with a tethered sun shield},
volume = {120},
journal = {Proceedings of the National Academy of Sciences of the United States of America},
}

@article{transparent_occulters,
  author = {Borgue, Olivia and Hein, Andreas M.},
  year = {2023},
  month = {01},
  pages = {308--318},
  title = {Transparent occulters: A nearly zero-radiation pressure sunshade to support climate change mitigation},
  volume = {203},
  journal = {Acta Astronautica},
}

@inproceedings{dimsun,
author = {Bandyopadhyay, Saptarshi and Bhamidipati, Sriramya and Silva, Maira Saboia da and Richardson, Mark and Hakuba, Maria and Lebsock, Matthew and Paranjape, Aditya and Nanjangud, Angadh and Jadhav, Tushar and Percival, Carl and Fishbein, Evan and Reager, John and Rahmani, Amir},
year = {2025},
month = {03},
pages = {1-15},
title = {Dimming the Sun (DimSun) using Controllable Swarm of Smallbody Regolith Particles},
}

@misc{ESA2023aluminium,
  author = {{European Space Agency}},
  title  = {To See the Universe in Aluminium},
  year   = {2023},
  note   = {\url{https://www.esa.int/Enabling_Support/Space_Engineering_Technology/To_see_the_Universe_in_aluminium} Accessed: 2026-06-12}
}

@misc{fairingstarship,
  author = {{eoPortal -- The Earth Observation Portal}},
  title  = {Starship of SpaceX},
  year   = {2025},
  note   = {\url{https://www.eoportal.org/other-space-activities/starship-of-spacex#starship-of-spacex} Accessed: 2026-06-12}
}

@article{90aluminium,
  author  = {Lindseth, I. and Bardal, A. and Spooren, R.},
  title   = {Reflectance measurements of aluminium surfaces using integrating spheres},
  journal = {Optics and Lasers in Engineering},
  year    = {1999},
  volume  = {32},
  number  = {5},
  pages   = {419--435},
  doi     = {10.1016/S0143-8166(00)00010-5}
}

@techreport{NASA_Materials2018,
  author      = {Finckenor, Miria M. and Tennyson, Robert C.},
  title       = {Materials for Spacecraft},
  institution = {NASA Marshall Space Flight Center},
  year        = {2018},
  number      = {NASA/TP-2016-219182}
}

@misc{alufoil,
  author = {{European Aluminium Foil Association}},
  title  = {Alufoil Production},
  year   = {2026},
  note   = {\url{https://www.alufoil.org/Alufoil-Production} Accessed: 2026-06-12}
}

@misc{5_6aluminium,
  author = {{Chalco Aluminum}},
  title  = {Classification of Aluminum Foil: A Complete Guide},
  year   = {2025},
  note   = {\url{https://www.chalcoaluminum.com/blog/aluminum-foil-classification-2505-lx/} Accessed: 2026-06-12}
}

@article{2althickness,
  author  = {Nie, Ning and Wang, Rui and Deng, Guanyu and Wang, Hui and Wang, Pengfei and Tieu, Anh and Li, Huijun and Su, Lihong},
  title   = {Fabrication of ultrathin Al foils by accumulative pack rolling: Influence of pack materials},
  journal = {Metals},
  year    = {2024},
  volume  = {14},
  pages   = {1262}
}

@misc{1micronAL,
  author = {{Fisher Scientific}},
  title  = {Aluminum Ultrathin foil, 1.0 micron thick, 99.995\% (metals basis), Thermo Scientific},
  year   = {2026},
  note   = {Catalog No. AA40684DJ; Supplier: Thermo Scientific Chemicals; Supplier No. 40684DJ}
}

@misc{nanoAL,
  author = {{American Elements}},
  title  = {Ultra Thin Aluminum Nanoscale Foil},
  year   = {2026},
  note   = {\url{https://www.americanelements.com/ultra-thin-aluminum-nanoscale-foil-7429-90-5} Accessed: 2026-06-12}
}

@article{ikaros,
  author  = {Tsuda, Y. and Mori, O. and Funase, R. and Sawada, H. and Yamamoto, T. and Saiki, T. and Endo, T. and Kawaguchi, J.},
  title   = {Flight status of {IKAROS} deep space solar sail demonstrator},
  journal = {Acta Astronautica},
  year    = {2011},
  volume  = {69},
  number  = {9},
  pages   = {833--840},
}

@article{nanosail,
  author  = {Spencer, David A. and Johnson, Les and Long, Alexandra C.},
  title   = {Solar sailing technology challenges},
  journal = {Aerospace Science and Technology},
  year    = {2019},
  volume  = {93},
  pages   = {105276},
}

@article{Lightsail2,
  author  = {Spencer, David A. and Betts, Bruce and Bellardo, John M. and Diaz, Alex and Plante, Barbara and Mansell, Justin R.},
  title   = {The {LightSail} 2 solar sailing technology demonstration},
  journal = {Advances in Space Research},
  year    = {2021},
  volume  = {67},
  number  = {9},
  pages   = {2878--2889},
  note    = {Solar Sailing: Concepts, Technology, and Missions II},
}

@article{revsolarsail,
  author  = {Gong, S. P. and Macdonald, Malcolm},
  title   = {Review on solar sail technology},
  journal = {Astrodynamics},
  year    = {2019},
  volume  = {3},
}

@article{lu2022conical,
  author  = {Lu, Lu and Dang, Xiangxin and Feng, Fan and Lv, Pengyu and Duan, Huiling},
  title   = {Conical Kresling origami and its applications to curvature and energy programming},
  journal = {Proceedings of the Royal Society A},
  year    = {2022},
  volume  = {478},
  number  = {2257},
  pages   = {20210712}
}

@article{misseroni2024origami,
  author  = {Misseroni, Diego and Pratapa, Phanisri P. and Liu, Ke and Kresling, Biruta and Chen, Yan and Daraio, Chiara and Paulino, Glaucio H.},
  title   = {Origami engineering},
  journal = {Nature Reviews Methods Primers},
  year    = {2024},
  volume  = {4},
  number  = {1},
  pages   = {40}
}

@misc{Starship,
  author = {{SpaceX}},
  title  = {Starship},
  year   = {2025},
  note   = {\url{https://www.spacex.com/vehicles/starship} Accessed: 2026-06-12}
}

@article{launchcapacity,
  author  = {Kurela, Michal and Nogues, Olivier and Alasa, Jean-Frédéric},
  title   = {Nondeterministic polynomial time algorithm for estimation of space launch base launch capacity},
  journal = {Acta Astronautica},
  year    = {2025},
  volume  = {234}
}

@article{Fuglesang,
  author  = {Fuglesang, Christer and García de Herreros Miciano, María},
  title   = {Realistic sunshade system at {L1} for global temperature control},
  journal = {Acta Astronautica},
  year    = {2021},
  volume  = {186},
  pages   = {269--279},
}

@article{sanchez,
  author  = {Sánchez, Joan-Pau and McInnes, Colin R.},
  title   = {Optimal Sunshade Configurations for Space-Based Geoengineering near the Sun-Earth {L1} Point},
  journal = {PLOS ONE},
  year    = {2015},
  volume  = {10},
  number  = {8},
}

@incollection{propellantless,
  author    = {Farres, Ariadna},
  title     = {Propellant-less systems},
  booktitle = {Next Generation CubeSats and SmallSats: Enabling Technologies, Missions, and Markets},
  editor    = {Branz, F. and Cappelletti, C. and Ricco, A. J. and Hines, J. W.},
  publisher = {Elsevier},
  year      = {2023},
  chapter   = {21},
  pages     = {519--541},
}

@misc{spectralon,
  author = {{Labsphere}},
  title  = {Space-Grade Spectralon\textsuperscript{\textregistered} Diffuse Reflectance Material},
  note   = {\url{https://www.labsphere.com/product/space-grade-spectralon-diffuse-reflectance-material/} Accessed: 2026-06-12}
}

@article{AL_values,
  author  = {Sznajder, Maciej and Seefeldt, Patric and Spr{\"o}witz, Tom and Renger, Thomas and Kang, J. H. and Bryant, R. and Wilkie, W.},
  title   = {Solar sail propulsion limitations due to hydrogen blistering},
  journal = {Advances in Space Research},
  year    = {2021},
  volume  = {67},
  number  = {9},
  pages   = {2655--2668},
}

@article{kidambi2020dynamics,
  author  = {Kidambi, N. and Wang, K. W.},
  title   = {Dynamics of Kresling origami deployment},
  journal = {Physical Review E},
  year    = {2020},
  volume  = {101},
  number  = {6},
  pages   = {063003}
}

@article{fang2017dynamics,
  author  = {Fang, Hongbin and Li, Suyi and Ji, Huimin and Wang, Kon-Wel},
  title   = {Dynamics of a bistable Miura-origami structure},
  journal = {Physical Review E},
  year    = {2017},
  volume  = {95},
  number  = {5},
  pages   = {052211}
}

@misc{acs3,
  author       = {Wilkie, Keats and Fernandez, Johnny},
  title        = {Advanced Composite Solar Sail System (ACS3) Mission Update},
  year         = {2023},
  institution  = {NASA Langley Research Center},
  note         = {Presented at the 6th International Symposium on Space Sailing, New York, NY, 5--9 June 2023; NASA Technical Reports Server Document 20230008378; \url{https://ntrs.nasa.gov/citations/20230008378} Accessed: 2026-06-12}
}

@article{designsolarsail,
  author  = {Zhao, Pengyuan and Wu, Chenchen and Li, Yangmin},
  title   = {Design and application of solar sailing: A review on key technologies},
  journal = {Chinese Journal of Aeronautics},
  year    = {2023},
  volume  = {36},
  number  = {5},
  pages   = {125--144},
}

@article{ghassaei2018fast,
  author  = {Ghassaei, Amanda and Demaine, Erik D. and Gershenfeld, Neil},
  title   = {Fast, interactive origami simulation using GPU computation},
  journal = {Origami},
  year    = {2018},
  volume  = {7},
  pages   = {1151--1166}
}

@inproceedings{ravichand2024origami,
  author    = {RaviChand, V. Sri Pavan and Kolathaya, Shishir and Balaji, K.},
  title     = {Origami for deployment of rigid panel array during space applications},
  booktitle = {2024 IEEE Space, Aerospace and Defence Conference (SPACE)},
  year      = {2024},
  pages     = {709--713},
  organization = {IEEE}
}

@article{yuan2022comparison,
  author  = {Yuan, Laohu and Song, Rui and Wang, Feng and Wei, Jianzheng and Liu, Jiafu},
  title   = {Comparison of two folded methods of solar sails},
  journal = {Aerospace},
  year    = {2022},
  volume  = {9},
  number  = {12},
  pages   = {746}
}

@inproceedings{gardsback2007design,
  author    = {G{\"a}rdsback, Mattias and Tibert, Gunnar and Izzo, Dario},
  title     = {Design considerations and deployment simulations of spinning space webs},
  booktitle = {48th AIAA/ASME/ASCE/AHS/ASC Structures, Structural Dynamics, and Materials Conference},
  year      = {2007},
  pages     = {1829}
}



%% \begin{thebibliography}{00}

%% For numbered reference style

%% \bibitem{lamport94}
%%  Leslie Lamport,
%%  \textit{\LaTeX: a document preparation system},
%%  Addison Wesley, Massachusetts,
%%  2nd edition,
%%  1994.

%% \end{thebibliography}

\newpage
\appendix

\section{Solar radiation pressure-induced forces}
\label{appendix:force_derivation}
In this appendix we derive the forces exerted by solar radiation that various geometries of sunshade would experience due to the effects of reflection, absorption and thermal re-emission. From our derivation of force, we can yield the sunshade-geometry specific momentum-transfer coefficient. By setting the appropriate conditions we can recover both our conical sunshade formulation (Eq.~\ref{eq:kappa_eff}) and McInnes' formulation (Eq.~\ref{eq:kappa_perp}).

The solar radiation pressure $P_{\mathrm{srp}}$ is the force per unit area exerted by sunlight on a surface. For a surface located at heliocentric distance $r_o-r_s$, it is given by
\begin{equation}\label{eq:Psrp}
    P_{\mathrm{srp}} = P_E \left(\frac{r_o}{r_o-r_s}\right)^2,
\end{equation}
where $P_E$ denotes the solar radiation pressure at 1 AU. Equivalently, $P_E = I_o/c$, where $I_o$ is the solar flux at 1 AU and $c$ is the speed of light.

The total force $\vec{F}$ exerted on a flat surface by solar radiation is the sum of the contributions due to specular reflection, diffuse reflection, absorption, and thermal re-emission. We write
\begin{equation*}
    \vec{F} = f_s \vec{F}_s + f_d \vec{F}_d + f_a \vec{F}_a + f_F \vec{F}_{\varepsilon_F} + f_B \vec{F}_{\varepsilon_B},
\end{equation*}
where $\vec{F}_s$, $\vec{F}_d$, and $\vec{F}_a$ denote the force contributions associated with specular reflection, diffuse reflection, and absorption, respectively, while $\vec{F}_{\varepsilon_F}$ and $\vec{F}_{\varepsilon_B}$ denote the Lambertian thermal re-emission contributions from the Sun-facing and Earth-facing sides of the surface.
The coefficients $f_s$, $f_d$, and $f_a$ are the fractions of incident radiation associated with specular reflection, diffuse reflection, and absorption, respectively, and satisfy $f_s + f_d + f_a = 1.$
The absorbed fraction is re-emitted thermally on the front and back sides, so that $f_a = f_F + f_B.$

Let $A$ be the surface area, let $\vec e_n$ be the unit normal to the surface, and let $\vec e_S$ be the unit vector pointing from the Sun to the object. Defining the inner product $\nu = \langle \vec e_n, \vec e_S \rangle$ between these unit vectors, the individual contributions can be written as
\begin{align*}
    \vec{F}_a &= P_{\mathrm{srp}} A |\nu|\, \vec e_S, \\
    \vec{F}_s &= -2 P_{\mathrm{srp}} A \nu^2 \vec e_n, \\
    \vec{F}_d &= P_{\mathrm{srp}} A |\nu| \left(\vec e_S - \frac{2}{3}\vec e_n\right), \\
    \vec{F}_{\varepsilon_F} &= -\frac{2}{3} P_{\mathrm{srp}} A |\nu|\, \vec e_n, \\
    \vec{F}_{\varepsilon_B} &= \frac{2}{3} P_{\mathrm{srp}} A |\nu|\, \vec e_n.
\end{align*}
All components of the total force involve the aforementioned inner-product, as this accounts for projected area of the surface with respect to the incoming solar radiation. For the absorption component, this force simply acts in the direction pointing to Earth. For the specular reflection component, the factor of two arises because the surface reverses the photon momentum component normal to the surface, doubling the normal momentum exchange, while the force is proportional to the surface normal ($\vec e_n$) because the tangential momentum component is unchanged. For the diffuse reflection component, the $\vec e_s$ term comes from transferring the incident photon momentum from the incoming Sun-line beam. Whereas the $-\frac{2}{3}\vec {e}_n$ term comes from Lambertian scattering: by Lambert’s cosine law, the average outgoing normal momentum produces a recoil opposite to the surface normal.
By the Stefan-Boltzmann law one can calculate the emitted flux of a material at temperature $T$ as
\begin{equation*}
    I = \varepsilon\sigma T^4,
\end{equation*}
with $\varepsilon$ the emissivity of the surface emitting radiation and $\sigma$ the Stefan-Boltzmann constant. Writing the emitted flux at the front (back) as $I_F$ ($I_B$). and assuming an emissivity at the front $\varepsilon_F$ and back $\varepsilon_B$ of the object we obtain from energy conservation
\begin{equation}\label{eqapp:flux_conservation}
    f_a I_o = I_F + I_B = (\varepsilon_F + \varepsilon_B)\sigma T^4.
\end{equation}
From \eqref{eqapp:flux_conservation} it follows that $\sigma T^4 = \frac{f_a I_o}{\varepsilon_F+\varepsilon_B}$ and hence one can write
\begin{align*}
    I_F & = \varepsilon_F \frac{f_a I_o}{\varepsilon_F+\varepsilon_B} = f_F I_o \Rightarrow f_F =  f_a\frac{\varepsilon_F}{\varepsilon_F+\varepsilon_B}, \\
    I_B & = \varepsilon_B \frac{f_a I_o}{\varepsilon_F+\varepsilon_B} = f_B I_o\Rightarrow f_B =  f_a\frac{\varepsilon_B}{\varepsilon_F+\varepsilon_B}.
\end{align*}
These expressions can be used to obtain the total force $\vec{F}$ in terms of $\varepsilon_F$ and $\varepsilon_B$, rather than in terms of $f_F$ and $f_B$. In the following subsections a flat surface perpendicular to the incoming light from the Sun, a tilted flat surface, and a cone will be considered.
\subsection{A flat surface perpendicular to the incoming light}
Assume a surface perpendicular to the incoming light, similar to the one considered in McInnes \cite{mcinnes}. For such an object one can write $\vec{e}_n  = -\vec{e}_S$ and $\vert \nu\vert = 1$. 
Hence, after substitution of $\vec{e}_S$, $f_F$ and $f_B$ one obtains
\begin{equation}\begin{split}
    \vec{F}_{srp} &= \biggl((f_s - 1) -2(f_s + \frac{1}{3}f_d) \\ &- \frac{2}{3}(1- f_s - f_d)\frac{\varepsilon_F - \varepsilon_B}{\varepsilon_F + \varepsilon_B} \biggr) P_{srp} A\vec{e}_n, \label{eq:F_srp_flat}
    \end{split}
\end{equation}
where $f_a = 1 - f_s - f_d$ was used. From the definition of the effective momentum-transfer coefficient $\kappa$ in Eq.\eqref{eq:kappa_def} and substituting in Eq.\eqref{eq:F_srp_flat} with respect to the Sun-Earth line ($\vec{e}_S$), we can recover the form of $\kappa$ demonstrated by McInnes \cite{mcinnes} as shown in Eq.~(\ref{eq:kappa_perp}) when $f_a=1-f_s$, $f_d=0$ and $\vec{e}_n=-\vec{e_s}$; consequently,
\begin{equation*}
    \kappa=\frac{1}{2}\left[1+f_s+\frac{2}{3}(1-f_s)\frac{\varepsilon_F-\varepsilon_B}{\varepsilon_F+\varepsilon_B}\right].
\end{equation*}
That is, the momentum-transfer coefficient for a flat object which exhibits specular reflection, absorption, and Lambertian thermal re-emission in the front and back.  
\subsection{A flat tilted surface}
More generally, we can take an example a tilted surface and calculate the forces acting. We will mainly be interested in the projection of the force on $\vec{e}_S$. This can be achieved by substituting $\vec{e}_n \rightarrow \nu \vec{e}_S$ (note the absence of the absolute values). We can write for the force components along $\vec{e}_S$
\begin{align*}
    \vec{F}_a & = P_{srp} ~A ~\vert\nu\vert \vec{e}_S,\\
    \vec{F}_s &= -2 P_{srp} A \nu^3\vec{e}_S,\\
    \vec{F}_d &= P_{srp} A \vert \nu\vert\left(1 - \frac{2}{3}\nu\right)\vec{e}_S,\\
    \vec{F}_{\varepsilon_F} &= -\frac{2}{3}P_{srp}A\vert\nu\vert\nu  \vec{e}_S, \\
    \vec{F}_{\varepsilon_B} &= \frac{2}{3}P_{srp}A\vert\nu\vert \nu  \vec{e}_S.
\end{align*}

\subsection{A cone}
Consider a hollow cone with with half-angle $\alpha$, whose height axis is along $\vec{e}_S$: its apex pointed at the sun and its (open) base pointed towards the Earth. For such a cone one can use its radial symmetry to note that any force component which is not along $\vec{e}_S$ gets cancelled out. Thus, one only needs to consider force components along $\vec{e}_S$.
The surface of the cone is an angled flat surface under angle $\alpha$. Equivalently, the angle between $\vec{e}_S$ and the normal unit vector $\vec{e}_n$ of this surface is given by $\theta = 90^\circ + \alpha$. As a result, one can write $\nu=\cos\theta=-\sin\alpha$.

In this work it is assumed that the solar radiation pressure $P_{srp}$ at each point of the cone along its height is approximately the same. Note however, that it is also straight-forward to derive expressions for a distance dependent $P_{srp}$, in case such an assumption fails. This is beyond the scope of this appendix.

For this particular case of a cone one can write for the force
\begin{align}
\begin{split}
\label{F_nodistdep}
    \vec{F}&=P_{srp} ~A \sin\alpha\biggl(f_a +2f_s \sin^2\alpha
    + f_d\left(1 + \frac{2}{3}\sin\alpha\right)\\ &+ \frac{2}{3}f_a \frac{\varepsilon_F -\varepsilon_B}{\varepsilon_F+\varepsilon_B}\sin\alpha\biggr)\vec{e}_S.
\end{split}
\end{align}
Here one can identify $A \sin\alpha$ as the surface of the base of the cone, and the effective surface that experiences solar flux. Under this construction it clearly is most advantageous to have $f_a = f_d = 0$ and minimize $\alpha$.

From Eq.\eqref{eq:kappa_def}, we can find the effective momentum-transfer coefficient for our conical sunshade by using Eq.\eqref{F_nodistdep} (where $A\sin\alpha=A_S$ is identified), we find \begin{equation}\begin{split}\label{eq:kappa_eff_append}
\kappa(\alpha)
&= \frac{1}{2}\biggl[
f_a+2 f_s \sin^2\alpha+f_d\left(1+\frac{2}{3}\sin\alpha\right)\\ &+\frac{2}{3}f_a\sin\alpha\frac{\varepsilon_F-\varepsilon_B}{\varepsilon_F+\varepsilon_B} \biggr],\end{split}
\end{equation} where Eq.\eqref{eq:kappa_eff_append} is our expression for the momentum transfer coefficient as shown in Eq.\eqref{eq:kappa_eff}.

\section{\label{Appendix:Reoptimisation}Analytic approximation thickness dependent cone position}
In this appendix an analytic approximation is derived for how the position $r_s$ depends on the thickness $t$ and the half-angle $\alpha$ of a cone made of a material with mass density $\rho$ and projected sunshade area $A_s$ in the Sun-Earth direction. To this end, one can equate Eq.\eqref{eq:1Ddynamics} to Eq.\eqref{eq:SRP}, and substitute $\epsilon=\frac{r_s}{r_o}$ and $M_s=\frac{A_s\rho t}{\sin\alpha}$, yielding
\begin{equation*}
    -\frac{GM_E}{r_o^2}(1-2\epsilon+\epsilon^2)-\frac{GM_o}{r_o^2}(-3\epsilon^3+3\epsilon^4-\epsilon^5) - \frac{2\kappa P_E \sin \alpha}{\rho t}\epsilon^{2} = 0.
\end{equation*}
Considering as a material Aluminium, as well as $\epsilon\approx\mathcal{O}(10^{-2})$, and the values of the thickness $t$ and angle $\alpha$ considered in this work, one obtains to non-trivial order the approximate equation
\begin{equation*}
    -A-B\epsilon^2 +C\epsilon^3=0,
\end{equation*}
 where $A=\frac{GM_E}{r_o^2}$, $B=\frac{2\kappa P_E\sin\alpha}{\rho t}$, and $C=3\frac{GM_o}{r_o^2}$. This equation can be solved exactly and allows one purely real solution 
\begin{align*}
    \epsilon &= \frac{\left(3\sqrt{3} \sqrt{27 A^2 C^4 + 4 A B^3 C^2} + 27 A C^2 + 2 B^3\right)^{1/3}}{2^{1/3} 3C}\\
    &+ \frac{2^{1/3} B^2}{3C (3 \sqrt{3} \sqrt{27 A^2 C^4 + 4 A B^3 C^2} + 27 A C^2 + 2 B^3)^{1/3}} + \frac{B}{3C}.
\end{align*}
Since several terms in this solution are of small magnitude, one can simplify the expression to leading order and obtain
\begin{equation}
    \epsilon(t,\alpha)\approx \epsilon_0 + \epsilon_1\frac{\kappa(\alpha)\sin\alpha}{t}+\epsilon_2\frac{\kappa^2(\alpha)\sin^2\alpha}{t^2},
\end{equation}
where $\epsilon_0 =\left(\frac{1}{3}\frac{M_e}{M_o}\right)^{1/3}$, $\epsilon_1=\frac{2}{9}\frac{P_E}{G M_o\rho }r_o^2$, and $\epsilon_2=4\frac{3^{1/3}}{3^{4}}\frac{P_E^2}{G^2M_E^{1/3}M_o^{5/3}\rho^2}r_o^4$.

\end{document}